\documentclass[11pt]{article}
\usepackage{subfigure}
\usepackage{float}
\usepackage{graphicx}
\usepackage{epsfig,epstopdf}
\usepackage{tikz}
\usepackage{enumerate}
\usepackage{bbm}
\usepackage{amsmath}
\usepackage{amsthm}
\usepackage{amssymb}
\usepackage{relsize,enumitem}
\usepackage{multirow,url}

\usepackage[top=0.9in,bottom=1.1in,left=1.05in,right=1.05in]{geometry}

\usepackage{pdflscape}

\newcommand\E{\ensuremath{\mathbb{E}}}

\newcommand\cM{\ensuremath{\mathfrak{M}}}
\newcommand\cL{\ensuremath{\mathfrak{L}}}

\newcommand\eps{\epsilon}

\DeclareMathOperator\sgn{sign}

\DeclareMathOperator\corr{corr}
\DeclareMathOperator\logit{logit}

\theoremstyle{remark}
\newtheorem{remark}{\textbf{Remark}}

\begin{document}

\title{Order Flows and Limit Order Book Resiliency on the Meso-Scale}
\author{Kyle Bechler\footnote{UCSB, Email: {kyle.bechler@gmail.com}}  \and Mike Ludkovski \footnote{ Corresponding Author: Department of Statistics and Applied Probability, University of California, Santa Barbara CA, 93106-3110 USA. Email: {ludkovski@pstat.ucsb.edu}}}
\date{June 29, 2017 \thanks{We thank Sebastian Jaimungal for many useful discussions and for help with the datasets. We are also grateful for feedback from the participants at the 2015 Stevanovich Conference on Market Microstructure and High Frequency Data (Chicago) and the 2016 Microstructure: Confronting Many Viewpoints Conference (Paris), where earlier versions of this work were presented.}}

\maketitle

\begin{abstract}
  We investigate the behavior of limit order books on the meso-scale motivated by order execution scheduling algorithms. To do so we carry out empirical analysis of the order flows from market and limit order submissions, aggregated from tick-by-tick data via volume-based bucketing, as well as various LOB depth and shape metrics. We document a nonlinear relationship between trade imbalance and price change, which however can be converted into a linear link by considering a weighted average of market and limit order flows. We also document a hockey-stick dependence between trade imbalance and one-sided limit order flows, highlighting numerous asymmetric effects between the active and passive sides of the LOB. To address the phenomenological features of price formation, book resilience, and scarce liquidity we apply a variety of statistical models to test for predictive power of different predictors. We show that on the meso-scale the limit order flows (as well as the relative addition/cancellation rates) carry the most predictive power. Another finding is  that the deeper LOB shape, rather than just the book imbalance, is more relevant on this timescale. The empirical results are based on analysis of six large-tick assets from Nasdaq.
\end{abstract}


\section{Introduction: the Meso-Scale}\label{sec:intro}
With the proliferation of the electronic trading, research on the fine time-scale of financial markets, exemplified by the microstructure of Limit Order Books (LOBs), continues to grow in scope and importance. The microstructure time-scale (on the order of milliseconds) works with the discrete events that correspond to the order messages submitted to the exchanges and hence captures the fundamental price formation and market behavior. It is often contrasted with the classical ``diffusion-limit'' framework (on the order of hours and days) that works with continuous-time processes and allows for infinitesimal trading. However, in between these two time-scales, there lies a third: the \emph{meso-scale}, capturing the asset behavior on the order of minutes.

The meso-scale is of key importance for the market participants engaged in market-making and order execution. Consider the proverbial trader who executes a large order (an action originally driven by a macro-scale analysis) by dividing it into smaller ``child'' orders executed over multiple ``slices''. She is concerned with execution costs comprised of instantaneous slippage and longer-term price impact. The first component is addressed via the \emph{routing} layer that optimizes individual trades, for example to avoid ``walking the book''. The second component is addressed by the \emph{scheduling} layer that determines the pace of execution and usually works over 2--10 minute intervals. This scheduling revolves around the expected evolution of the LOB through time, requiring aggregation of the information from the static LOB snapshots and lifting it to the meso-scale.

The dynamics of the book on the meso-scale, and the associated consumption and provision of liquidity drive the concept of resilience, also known as the transient and permanent price impacts, that underlies much of the literature on algorithmic trading and order execution. On the tick-by-tick scale, the effect of an order is {mechanical} and can be easily described using  observable LOB characteristics, namely the resting limit volume at various levels, summarized via LOB depth and shape. In particular, for vast proportion of executions direct price impact is zero, as only a portion of the respective top queue is consumed. However, on a longer time-scale (as implemented in order splitting algorithms that typically unfold over several hours), the enormous volume of  the order messages obscures the link between price evolution and individual orders, and renders static liquidity measures irrelevant. Consequently, proper calibration of the HFT models ought to rely on meso-scopic/dynamic, rather than microscopic/static LOB metrics.

To measure liquidity on the mesoscale it is necessary to consider aggregated metrics about types/volumes of orders submitted. A natural quantity of interest therefore becomes the \emph{order flows} which can be viewed as the dynamic version of instantaneous LOB characteristics. Statistical analysis of the order flows is the main focus of this article. We analyze the mesoscopic links between market and limit order flows, and between order flows, prices and static LOB metrics. Our main goal is a phenomenological description of these relationships, so as to a provide a data-driven perspective for LOB behavior. For empirical work we utilize a cross-section of 6 liquid, large-tick Nasdaq tickers spread across 2 calendar years.

 Based on our analysis, we report several key findings. First, we document an S-shaped nonlinear relationship between mid-price change $\Delta P$ and market order flow $VM$. In tandem, we also record a strong linear relationship between $\Delta P$ and Net Liquidity flow, which is a weighted average of $VM$ and  limit order flow $VL$. These empirical results offer an important correction to the commonly assumed framework of linear price impact: to obtain linearity one must take into account touch limit orders, weighted appropriately. Ignoring limit flow dramatically weakens the explanatory power of the models, in fact our findings suggest that limit flows are \emph{at least, if not more} statistically significant than the market trades.

 Second, we use statistical tools to identify key predictors for price formation on the meso-scale. We find that the explanatory power of the top-level LOB queue depths is rather weak. This is primarily due to the rapid fluctuations in those quantities, which do not allow for easy time-aggregation. Instead, liquidity measures that look deeper into the book are more relevant for quantifying liquidity. At the same time,  we find that the LOB metrics offer an important \emph{modulation} to the effect of order flows and do add a non-negligible predictive power. Third, we disentangle price trend (defined in terms of overall bid/ask-side pressure captured by market order flow $VM$) from liquidity (primarily captured by limit order flows $VL$). This allows us to pinpoint the meaning of \emph{scarce liquidity}. We carry out further statistical analysis in that direction, which is of interest for adaptive order scheduling that can respond to varying expected price impact.

The meso-scale is intrinsically defined via aggregating the raw LOB data into slices, and the question of aggregation method looms large. The micro-scale is naturally discrete-event-based. On the macro-scale, the high degree of averaging lends itself naturally to a (calendar) time-based limit.  Here we propose that \emph{volume-based} aggregation is most appropriate for the meso-scale. Thus, our analysis is based on
working with LOB buckets, with  each bucket containing the same amount of executed (market) volume. As we show, this yields a desired stabilizing effect on the LOB data and removes some known artifacts. It also quantifies what we mean by the meso-scale; informally it can be characterized as dealing with ``dozens of trades''; indeed our buckets contain 10--100 market executions, accompanied by 100--1000 limit order events. At that intermediate time-scale, the aforementioned event-based analysis is less tractable for capturing the dynamic quantities that come to the fore; while time-based analysis must contend with huge swings in market activity that are caused by extreme clustering of LOB events.

As far as we are aware, this is the first academic study of volume-aggregated intra-day LOB data.
Several previous studies used fixed-interval time-aggregation, such as 15-min slices in \cite{HasbrouckSeppi01} and 5-min slices in \cite{cartea2015incorporating}, \cite{HardleHautsch12} and \cite{benzaquen2016dissecting}. The closest study to ours is by Cont et al.~\cite{cont2013price} who also investigated the relationship between order flows, price change and book depth, and reached some similar conclusions. However, that study was limited to working only with top-level book information, and most crucially \emph{postulated} the expected (linear) link between the different variables. In contrast, we employ non-parametric tools to obtain data-driven findings about variable importance, empirical non-linearities, et cetera. Moreover, we also investigate the asymmetric features of the order flows which are obscured by the total netting carried out in \cite{cont2013price}.

Our investigation offers a statistical counterpoint to the research on optimal execution which has put forth a plethora of different \emph{theoretical} price impact models. From the empirical point of view, there is already a growing literature on tick-by-tick data, see \cite{cont2013price,donnelly2014ambiguity,Farmer,eisler2012price,Toth12,Toth15}. These are complemented by  LOB \emph{simulators} that aim to capture short-term behavior and price impacts. The most popular framework views the LOB as a collection of interacting queues \cite{HautschHuang12,moallemi2014short,lipton2013trade,Toke15}. For example, Huang et al~\cite{huang2014simulating} model limit order, cancellation, and market order arrival rates as a function of the queue sizes. Unfortunately such models are much less tractable on the meso-scale, so few analytic results are so far available. An alternative way to model the mesoscopic price impact is to infer the impulse response function, known as the propagator, of an \emph{individual} order/trade. The aggregate effect on the mid-price is then viewed as a linear superposition of individual impacts, see \cite{eisler2012price,bouchaud2004fluctuations}. However, such event-by-event analysis largely excludes capturing the statistical effects of observable covariates which ``dress'' the evolution of the  respective dynamical system describing the LOB. Yet another micro-scale analysis which also leads to important meso-scale implications is the emerging strand on long-memory effects of micro-structural phenomena \cite{RosenbaumMasaaki16}, as well as time-series approaches to the functional evolution of the entire LOB shape \cite{HardleHautsch12}.

The remainder of the paper is structured as follows. In Section \ref{sec:Liq} we discuss LOB evolution in the context of static and dynamic measures of liquidity and price impact, as well our volume-bucketing method.
Sections \ref{sec:results}-\ref{sec:liquidity} contain our main statistical analysis regarding the meaning of price trend, liquidity, and scarce liquidity. Section \ref{sec:ts} takes a different tack, addressing the time-series properties of order flows. Finally, Section \ref{sec:conclude} discusses ramifications from the previous sections and offers an outlook for further work.

\section{Limit Order Books: Price Impact, Order Flows and Liquidity}\label{sec:Liq}
	
	\subsection{Limit Order Book}\label{sec:lob}
Electronic trading marketplaces match liquidity providers and consumers via the limit order book (LOB). Participants asynchronously submit trading messages which are aggregated by the exchange into the LOB.
The two base classes of trades are market orders and limit orders. Market orders, henceforth called \emph{trades},  indicate actual transactions taking place and are denoted as $\cM := \{ (T^M_i, O^M_i) \}$, where $T^M_i$ are the execution times, and $O^M_i$ are corresponding execution volumes.  $O^M_i$'s are signed, with positive indicating a buy order and negative a sell order. Limit orders, henceforth \emph{orders}, are $\cL:= \{ (T^L_i, O^L_i, S^L_i) \}$ where $T^L_i$ are the message time stamps, $O^L_i$ are (signed) order volumes and $S^L_i$ is the limit order price. Positive values of $O^L_i$ indicate submission of a new limit order, while negative $O^L_i$ indicates a {cancellation}. Limit orders reside in the LOB until executed by an incoming market order, or being cancelled by the participant who initially placed the order. We refer to \cite{GouldSurvey13,HFT} for more general introduction on the functioning of LOBs, and other types of orders possible.
	
At any given moment $t$, the LOB lists all the resting limit orders, represented as a vector $(p^j_i(t), v^j_i(t))$, listing the volume $v_i(t) \ge 0$ of resting limit orders at price $p_i(t)$.  The superscript $j \in A,B$ denotes the Ask and Bid sides, respectively.  The subscript $i=1,2,\ldots$ indexes the price level which is discretized in terms of the tick size $\Delta p$, typically 1 cent for US equities.  Indexing is done  separately for each side of the LOB, starting from the best-bid and best-ask levels and moving consecutively. Thus, $p^A_1(t)$ is the best-ask (at the touch) price, $p^A_2(t) = p^A_1(t) + \Delta p$ is the second price level, etc. Some queues can be empty, $v^j_i(t) = 0$. However the top queues  $v^j_1(t) > 0$ are always strictly positive, and after netting executable limit orders, the best-ask price is at least one tick above the best-bid $p^A_1(t) \ge p^B_1(t) +\Delta p$. Based on the bid/ask prices, we have the midprice $P$ and the spread $S$:
	\begin{align}
	P(t) := \frac{ p^{A}_1(t) + p^{B}_1(t)}{2}, \qquad Spr(t) := p^{A}_1(t) - p^B_1(t).
	\end{align}

\subsection{Static LOB Snapshots and Price Impact}
\label{sec:static}
	
A starting point for understanding LOB liquidity is the information contained in a static snapshot of the LOB.
The most basic measure of liquidity is the spread $Spr(t)$ which dictates the costs of a round-trip trade that is fundamental to market-making. However, for liquid US equities the spread may not be useful, as it is not sensitive to changing market conditions. In particular for ``large tick'' assets, the bid-ask spread is almost always equal to one tick.  This is the case for all the assets in our sample, where $Spr(t)$ is essentially always 1 or 2 ticks -- and the 2-tick spreads are short-lived: after opening up the spread quickly closes as an aggressive limit order is posted inside the spread.

Another commonly used measure is the volume-at-the-touch $v^{j}_1(t)$, i.e.~queue lengths at the best-offer and best-bid price levels. Over very short time horizons, $v_1$ indicates the immediate liquidity available for incoming market orders, and hence drives the direction and likelihood of a mid-price change. Recall that $P$ changes by a half-tick when $v^j_1$ hits zero, whereupon $v^j_1$ is reset to the posted volume at the new top LOB level.
A related quantity that measures the relative ``strength'' of each side of the book is the Book Imbalance
	\begin{equation}\label{BI}
	BI := \frac{v^A_1 - v^B_1}{v^A_1 + v^B_1} \in (-1,1).
	\end{equation}
The book imbalance $BI$
	has been shown to be predictive of the next order side and resulting price move \cite{donnelly2014ambiguity, cont2013price,JaimungalDonnelly15}. Consequently, at the slicing/routing level of the execution process, $BI$ is often used as an indicator of when to employ limit orders and when to cross the spread and place a market order  \cite{stoikov2012optimal,lipton2013trade}. In a similar vein, \cite{huang2014simulating,Toke15,LehalleMounjid} document a link between $BI$ and
order arrival rates. 

The top queues offer only a ``tip of the iceberg'' summary for the overall LOB shape. The direct procedure to describe LOB shape employs units of shares and measures cumulative \emph{depth} over the first $k$ levels
\begin{align}
D^{j}_k := \sum_{i=1}^k v^j_i, \qquad j=A,B, \quad k=1,2,\ldots.
\end{align}
Thus, $D^j_1 = v^j_1$ is the top-level depth, $D_2^j = v^j_1 + v^j_2$ is the total volume posted at the top two levels and so on. A complementary inverse procedure measures \emph{execution cost} in units of ticks. Execution cost quantifies the slippage from immediately making a market trade, converting depth (shares) into monetary terms.
Fix a quantity $N$ of shares and let
$\bar{i}^j(N) :=\max \{k: D^j_k(t) < N\}$. Then
	\begin{equation}\label{PI}
	PI^j_N(t) := \frac{1}{N} \left\{\sum_{i=1}^{\bar{i}^j(N)} v^j_i(t) \left(p^j_i(t)-P(t)\right)+( N-D^j_{\bar{i}^j(N)}(t) ) \left(p^j_{\bar{i}^j(N)+1}(t)-P(t) \right)\right\}.
	\end{equation}
Thus, $PI^j_N$ blends information about the shape of the LOB by computing weighted average cost (relative to the mid-price) per share of immediately executing $N$ shares in the $j$-th direction. The map $N \mapsto PI^j_N(t)$ then summarizes the shape of the book in terms of executing different order sizes. Intuitively, $PI \propto 1/D$ is inversely proportional to depth.

 \begin{remark}
 We stress that $PI_N(t)$ is a theoretical construct, based on the counterfactual that an immediate market order is placed exactly at $t$. In practice the vast majority of market trades have zero impact, consuming less than the standing liquidity at the first level. Moreover, latency delays make it difficult to realize the ``observed'' $PI$ in real-life trades. 
 \end{remark}

\begin{figure}[htb]
		\centering 
\begin{tabular}{lr}
		\includegraphics[height=2.3in,trim=0.25in 0.2in 0.25in 0.2in]{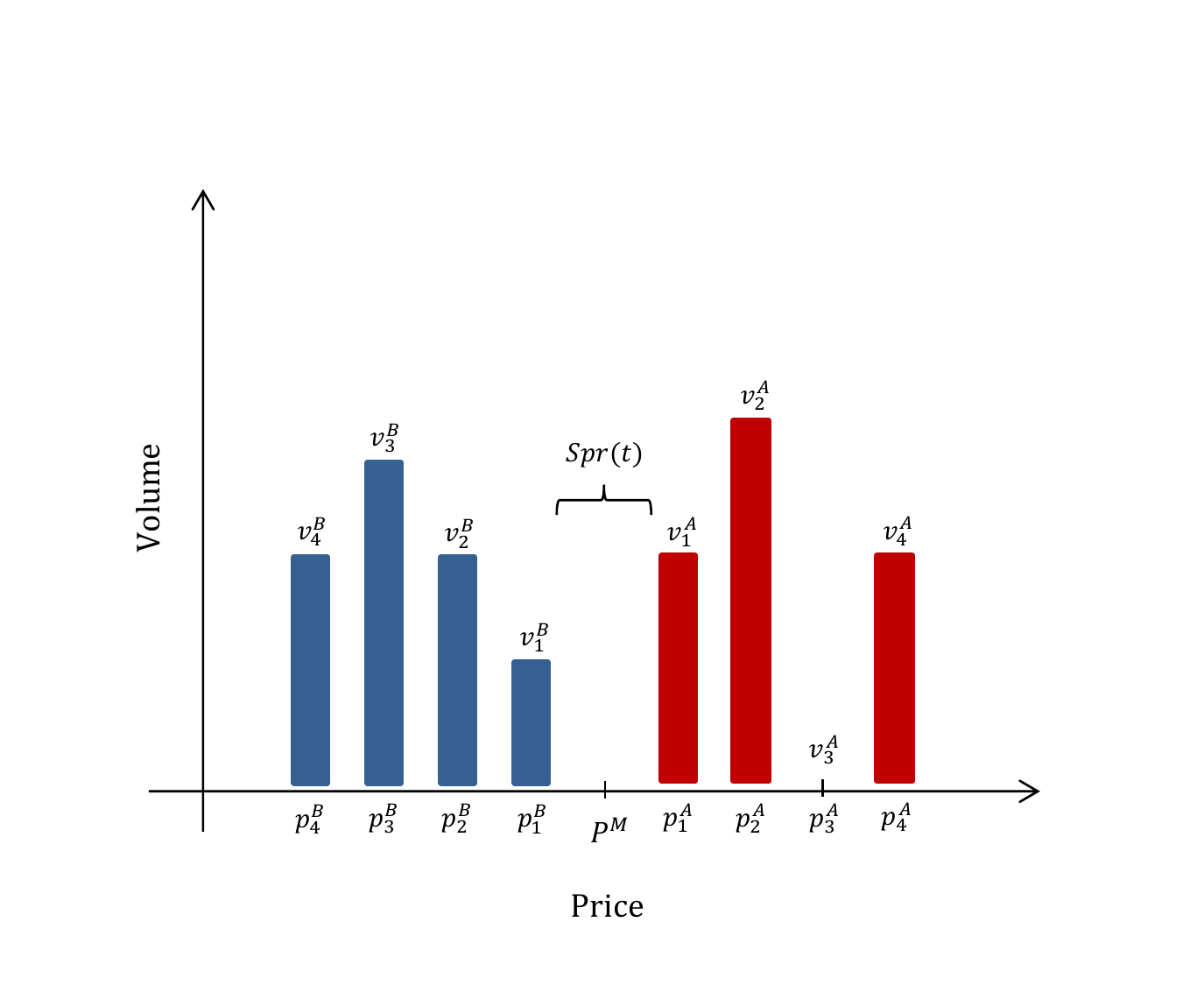} &
		\includegraphics[height=2.2in,trim=0in 0.2in 0in 0.3in]{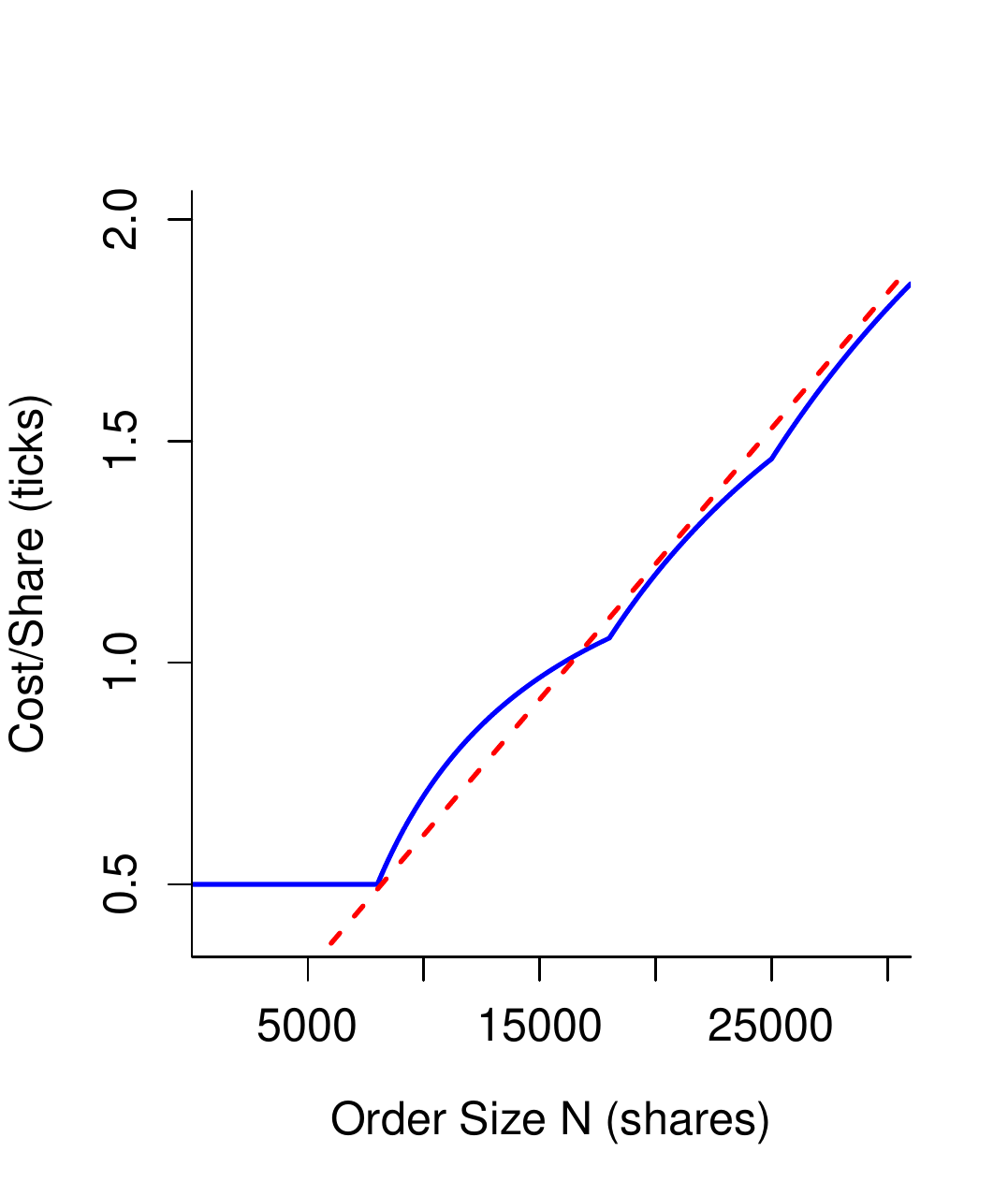}  \end{tabular}
		\caption{\emph{Left:} Stylized limit order book. In this illustration the spread is two ticks $Spr(t) = 2\Delta p$ and book imbalance is positive $v^A_1 > v^B_1 \Rightarrow BI(t) > 0$.
\emph{Right:} Measuring execution cost in a hypothetical limit order book. The underlying queue lengths are taken to be $(v_1, v_2, v_3, v_4) = (8000, 10000, 7000, 15000)$. We plot $n \mapsto PI_n$ (solid) and the fitted line $n \mapsto \hat{S} \cdot n$  based on the impact slope $\hat{S}$ (dashed).
			\label{fig:PIslope}}
	\end{figure}

The right panel of Figure~\ref{fig:PIslope} illustrates the above computations for a hypothetical book with $v_1 = 8000, v_2 = 10000, v_3  = 7000, v_4 = 15000$. The solid curve shows $N \mapsto PI_N$. Note that \eqref{PI} implies that for $N \le v_1$, $PI_N = Spr/2$ is half the spread; for larger order sizes, $PI_N$ has a nonlinear shape due to the $N$-dependent averaging weights. For example, at $N=30,000$, we obtain $PI_N = 1.8$ ticks, which averages the cost of 0.5 ticks for the first 8000 shares, 1.5 ticks for the next 10K shares, 2.5 ticks for the 7K thereafter and 3.5 ticks for the last 5K shares.

 One can observe from the plot that $N \mapsto PI_N$ is asymptotically linear, which would be consistent with the book having an asymptotic ``rectangular'' shape with $\bar{v}$ volume per level. (Namely, $\bar{v} \simeq N/(2 PI_N)$ which for the computation above yields $\bar{v} = 8333$ taking $N=30K$.)  Motivated by this idea, \cite{cartea2015incorporating} have proposed to compute an LOB slope $S^j$ which is obtained from a linear regression model
	\begin{equation}\label{slope}
	PI^j_n= S^j  \cdot n +\epsilon, \quad n=1,2,\ldots,N,
	\end{equation}
estimated using the standard least-squares fit. The slope coefficient $S^j$ can be interpreted as a linearized price impact (on the $j$-th side of the book) per share and allows to  imply a single liquidity metric from the collection $(PI^j_n)_n$. In Figure~\ref{fig:PIslope} the estimated impact slope is $\hat{S} = 0.0611$ ticks/1000 shares, see the fitted dashed line $\hat{S} \cdot n$. $\hat{S}$ can be inverted to map into ``averaged'' book depth $\bar{v} = 0.5/\hat{S} = 8176$.
An advantage of the metric $S$ is the ability to average $v_k$'s over multiple levels and incorporate the spread $Spr$, which makes it less sensitive to short-term fluctuations that dramatically lower the statistical usefulness of depth measures $D^j_k$.

	\subsection{Order Flows and LOB Evolution}\label{sec:lob-evolution}

\begin{figure}[htb!]
		\centering
		\includegraphics[width=0.85\textwidth, trim=0.1in 0.1in 0.1in 0.2in]{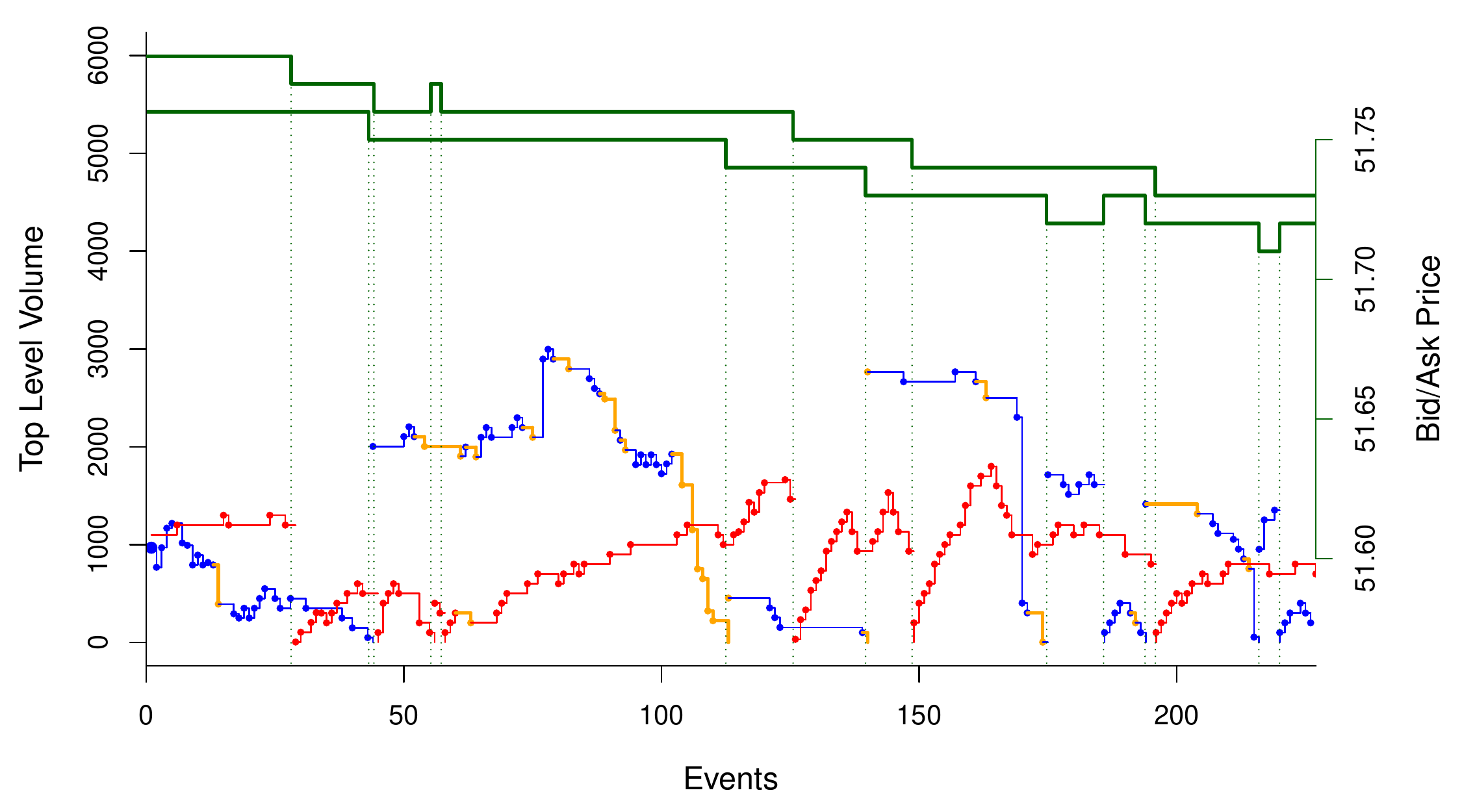}
		\caption{Best bid/ask queue levels $v_1^j(t)$ for TEVA (bottom curves) along with bid/ask price $p_1^j(t)$ (top). Event-by-event data taken from a $90$ second window beginning at 2:30pm on 2/18/2011.   Limit orders are in red (ask-side) and blue (bid), market executions in orange. All orders are plotted event-by-event (in particular without re-combining split market orders), so that the mechanics and sequence of order arrivals are more clear. \label{fig:lob-dynamics}}
	\end{figure}

To visualize the meso-scopic behavior of the book, Figure~\ref{fig:lob-dynamics} shows an event-by-event summary of an LOB for a representative 90-second period. The main plot at the bottom shows the top-level queue lengths $v^j_1$ driven by the market orders (in orange) and limit orders at-the-touch. At the top  of Figure~\ref{fig:lob-dynamics} we also track the contemporaneous effect on the mid-price, with dotted vertical lines marking a change in bid- or ask-price.

 The Figure indicates that following a market execution the LOB response varies widely, at times bouncing back through fresh posted volume, while at other times falling through and fading. We observe several ``regimes'': episodes of strong resilience when market orders are counteracted with added limit orders (so that $v^j_1(t)$ stays roughly constant over time), and other periods where market orders are accompanied primarily by limit order cancellations, creating a strong negative trend in $v^j_1(t)$, known as an LOB fade.  This happens on the left of Figure~\ref{fig:lob-dynamics}, where the arriving market flow is heavily tilted towards sell orders $TI < 0$ and net limit flow to the bid-side touch is clearly negative. The result is a rapid drop in mid-price, creating \emph{scarce} liquidity for sellers and increasing their respective execution costs. A milder form of this is when few new limit orders are added at the touch, so that the book is not replenished fast enough and ``retreats'' along with executed trades.
 Our goal below is to find meso-scopic predictors that explain whether market orders are likely to be met with strong resilience via new limit orders, or minimal liquidity provision or even net cancellations.

	\subsection{Volume Bucketing}\label{sec:bucketing}
	
	To move from the micro- to meso-scale we divide the trading day into \emph{buckets} that are used for aggregation. Buckets are defined in terms of their start- and end- time-stamps $\{\tau_k\}$, with the $k$-th bucket consisting of all messages entered during $(\tau_k, \tau_{k+1}]$ (special treatment possibly accorded to the first and last message, see below). Slice times $\tau_{k+1}$ are taken to be the time-stamp of the final trade included in bucket $k$.
The resulting bucketed quantities are:
	\begin{align}\label{VMeq}
	VM^{A,B}_k & := \sum_{i: \tau_k < T^M_i \le \tau_{k+1}} |O^M_i| 1_{\{O^M_i \gtrless 0\}}; \\ 
	VL^{j,(1)}_k & := \sum_{i: \tau_k < T^L_i \le \tau_{k+1}} O^L_i 1_{ \{S^L_i = p^j_1( T^L_i)\}}, \qquad j=A,B. \label{VLeq}
	\end{align}
Thus, $VM^j$ are the respective aggregated volumes of executed market buys and sells, and $VL^{j,(1)}$ are the volumes of limit orders at the touch. Note that $VL^{j,(1)}$ tracks the mid-price throughout the bucket, aggregating the respective top-level (at the time of message submission) limit orders. Because the tickers we consider are highly liquid in the sequel we focus exclusively on top-level limit orders $VL \equiv VL^{(1)}$. Recall that  market orders capture the consumption of liquidity by aggressive orders and decrease top-level depth $v^j_1$,  while limit orders can either augment $v^j_1$ (liquidity provision) or decrease it (order cancellations). Consequently, $VM^j \ge 0$ is non-negative by construction, whereas $VL^j$ can be of either sign depending on the net limit additions/cancellations.

Given a slice $(\tau_k, \tau_{k+1})$ we define the corresponding price change as
\begin{align}
  \Delta P_k = P(\tau_{k+1}) - P(\tau_k).
\end{align}
Based on the buckets we also define the discrete snapshots of LOB statistics through indexing by $k$; for example the book imbalance $BI_{k} \equiv BI(\tau_k)$ or the depth $D_k \equiv D(\tau_k)$. As analogue to the static $BI$, we also define the normalized \emph{trade imbalance}
	\begin{equation}\label{def:TI}
	TI_k :=\frac{VM^B_k-VM^A_k}{V_k},
	\end{equation}
where $V_k=VM^A_k + VM^B_k$ is the total executed volume in the bucket. By definition, $TI_k \in \{-1,1\}$; if $TI_k = 1$ then all executed trades in the bucket were `Buys', and if $TI_k = -1$, all executed trades were Sells.	Trade imbalance captures the aggregate supply and demand for the asset and is often used as the basis for measuring price impact \cite{bouchaud2008markets,cartea2015incorporating}.

We propose to bucket in terms of \emph{executed market volume}, namely keeping $V_k \equiv V$ fixed across buckets. Technically, this requires that market orders are sometimes split in two as one bucket ``fills'' up and the next begins. In our main analysis we take $V \simeq \{0.25\%, 1\%, 2\%\}$ of average daily volume (ADV, based on the considered time period of 10:00am--3:45pm) for the respective ticker. This means that \emph{on average} there are about 400 (resp.~100, 50) buckets per day; actual number of buckets can be 30\%--500\% of the average due to fluctuating levels of market activity.

Note that slicing is done in terms of executed volume;  volume of limit orders $VL^A_k + VL^B_k$ remains random across buckets. The reason for this is that even  though market orders account for only $2-4\%$ of total trades, they indicate actual transactions taking place and hence ultimately drive traders' P\&L. Thus, due to their intrinsic nature of ``putting money on the table'', they are typically viewed as influential by other participants and carry the most information for defining the ``business-time'' for the ticker.

Most related studies have used time-based aggregation. For instance, \cite{cont2013price} used 10-second slices, \cite{benzaquen2016dissecting,cartea2015incorporating,HardleHautsch12} used 5-min slices. Nevertheless,
using volume buckets has several attractive properties for our purposes. First, slicing by trade volume rather than time yields consistency in information across buckets. Under clock-time bucketing, one can easily appreciate the difficulty of comparing buckets with dramatically different activity levels (in the extreme, a time-based bucket could be entirely empty). Second, working in volume-time also reduces the intra-day seasonality effects, such as volume and volatility clustering, and improves statistical properties (such as Gaussianity) of the $(VL_k, VM_k)$ time-series. Third, comparison of limit order activity across buckets is naturally normalized when measuring in (trade) volume time, i.e.~the stable quantity is not $VL_k$ per se, but rather the ratio $VL_k/V_k$ (or $VL_k/VM_k$). For example, it is intuitive that limit additions to the touch depend more on quantity of market trades than passing minutes. Fourth, our method avoids the difficult challenge of estimating trade volume effects that are present in trade-by-trade aggregation.

\subsection{Datasets}\label{sec:data}
For our statistical analysis
we use Nasdaq ITCH TotalView data which contains ``Level-2'' information on order book events.
Specifically, we have access to direction and size of market executions, as well as limit order additions, modifications and cancellations for $i \le 30$ levels (ticks) into the current book. To avoid synchronization issues, we work only with Nasdaq-based messages, leaving out information from other exchanges. Furthermore, since erratic LOB behavior is often seen near the market open and close, we consider only activity between 10:00am and 3:45pm. Other pre-processing included removing all executions against hidden orders (less than $10\%$ of executed volume) and aggregating executed trades. Market orders are often matched against several smaller limit orders; we re-created the size of the original market order by aggregating orders that were consecutive, in the same direction, and with identical time stamps.

To give a broad slice of market activity, we use a total of six  different tickers, covering two  distinct time-periods. Thus, we used data on MSFT, TEVA and BBBY from the first 100 days of 2011, and data on INTC, ORCL and NTAP from the last 100 days of 2013. All assets are categorized as large cap stocks, with market caps ranging from about 12B USD for BBBY, and up to 240B for MSFT.
 Tables \ref{tab1}-\ref{tab2} provide summary info for each stock.  We note that all stocks are highly liquid, with several thousand trades per day, and hundreds of thousands of limit orders. Moreover, the analyzed LOBs are quite deep, with average depth of 3--10 average event size (AES). As a result, most of the economically relevant information can be gleaned from the top-level events. These still  correspond to 50K--100K touch limit orders per day, and respective nominal order volume $\sum_k |VL_k| \simeq$ 5M--100M shares.

	\begin{table}[htb]
		\begin{center}$$
			\begin{array}{rrrrrrr}
			\hline
			& \multicolumn{2}{ c }{ \text{MSFT}} & \multicolumn{2}{ c }{ \text{TEVA}} &\multicolumn{2}{ c }{ \text{BBBY}} \\ 	\hline
			& \text{mean } & \text{(stdev)}& \text{mean}& \text{(stdev)} & \text{mean} & \text{(stdev)}\\ \hline\hline
\text{Mean Price (USD) } P & 26.29 & (1.25) & 50.34 & (2.73) & 50.51 & (3.64)\\
\text{Mean Abs Price Change (ticks)} & 1.80  & (1.46) &  3.33 & (3.07) & 3.67 & (3.27) \\ 
\text{Ave Daily Vol (shares) } VM & 10.0M & (4.61M) & 1.21M & (1.06M) & 0.64M & (0.30M) \\
\text{Num of Daily MO's (events)} & 5340 & (2173) & 5455 & (2484) & 4281 & (1483) \\		            												\text{Ave Event Size (shares)} & 1643 & (3612) & 293 & (596)  & 192 & (321) \\
\text{Ave Limit volume (shares) } VL & 116.4M & (42.6M) & 12.04M & (5.90M) & 6.60M & (2.58M)  \\
\text{Num of Daily LO's (events)} & 292K & (92K) & 110K & (45K) & 77K & (31K) \\
\hline  
			\text{Spread  (ticks) } Spr & 1.01 & (.067) & 1.13 & (.362) & 1.29 & (.599)\\  
			\text{Mean depth at Level 1 } D_1 & 25.5K & (27.9K) & 1.33K & (2.62K) & 0.63K & (1.09K) \\
            \text{Mean depth at Levels 1+2 } D_2 & 59.2K & (29.9K) & 2.80K & (7.4K) & 1.24K & (0.97K) \\  
       \text{Ave  Slope (ticks/000's shares) } S & .0274 & (.00976) & 0.691 & (0.305) & 1.37 & (0.663) \\ 
	\hline		\end{array}$$
		\end{center}
		\caption{Summary statistics (mean/standard deviation) for MSFT (Microsoft), BBBY (Bed Bath \& Beyond) and TEVA (Teva Pharmaceuticals) for the 100 trading days from $1/2/2011$ to $5/25/2011$. Average price change is per 1\%-ADV volume buckets.
			\label{tab1}}
	\end{table}

	\begin{table}[htb]
		\begin{center}$$
			\begin{array}{rrrrrrr}
			\hline
			& \multicolumn{2}{ c }{ \text{INTC}} & \multicolumn{2}{ c }{ \text{ORCL}} &\multicolumn{2}{ c }{ \text{NTAP}} \\ 	\hline
			& \text{mean } & \text{(stdev)}& \text{mean}& \text{(stdev)} & \text{mean} & \text{(stdev)}\\ \hline\hline
			\text{Mean Price (USD) } P & 23.66 & (0.98) & 33.84 & (1.28) & 41.26& (1.35)\\
            \text{Mean Abs Price Change (ticks)} & 1.31  & (1.10) &  2.11 & (1.79) & 2.71 & (2.27) \\
			\text{Ave Daily Vol (shares) } VM & 3.58M & (1.18M) & 1.84M & (0.80M) & 0.71M & (0.31M) \\  
            \text{Num of Daily MO's (events) } & 2058 & (609) & 2880 & (1192) & 2171 & (935) \\					
            \text{Ave Event Size (shares)} & 1740 & (3819) & 837 & (1257)  & 252 & (287) \\
            \text{Ave Limit Volume (shares) } VL  & 99.62M & (33.42M) & 40.30M & (15.96M) & 15.80M & (5.26M)  \\
            \text{Num of  Daily LO's (events)} & 263K & (77K) & 190K & (68K) & 136K & (46K) \\
\hline  
			\text{Spread (ticks) } Spr & 1.01 & (0.13) & 1.03 & (0.17) & 1.08 & (0.29)\\   
			\text{Mean depth at Level 1 } D_1 & 13.9K & (10.0K) & 3.9K & (2.8K) & 1.24K & (1.13K) \\  
			\text{Mean depth at Levels 1+2 } D_2 & 31.9K & (17.9K) & 8.5K & (4.8K) & 2.69K & (1.28K) \\  
            \text{Ave Slope (ticks/000's shares) } S & .0367 & (.0158) & 0.136 & (.0659) & 0.432 & (0.201) \\		\hline 
			\end{array}$$
		\end{center}
		\caption{Summary statistics (mean/standard deviation) for INTC (Intel), ORCL (Oracle), and NTAP (Net Appliances) for the  100 trading days from $8/9/2013$ to $12/31/2013$.
			\label{tab2}}
	\end{table}

\begin{figure}[ht]
		\centering
		\begin{tabular}{ccc} \hspace*{-0.16in}
			\includegraphics[height=2in,trim=0in 0.2in 0in 0.3in]{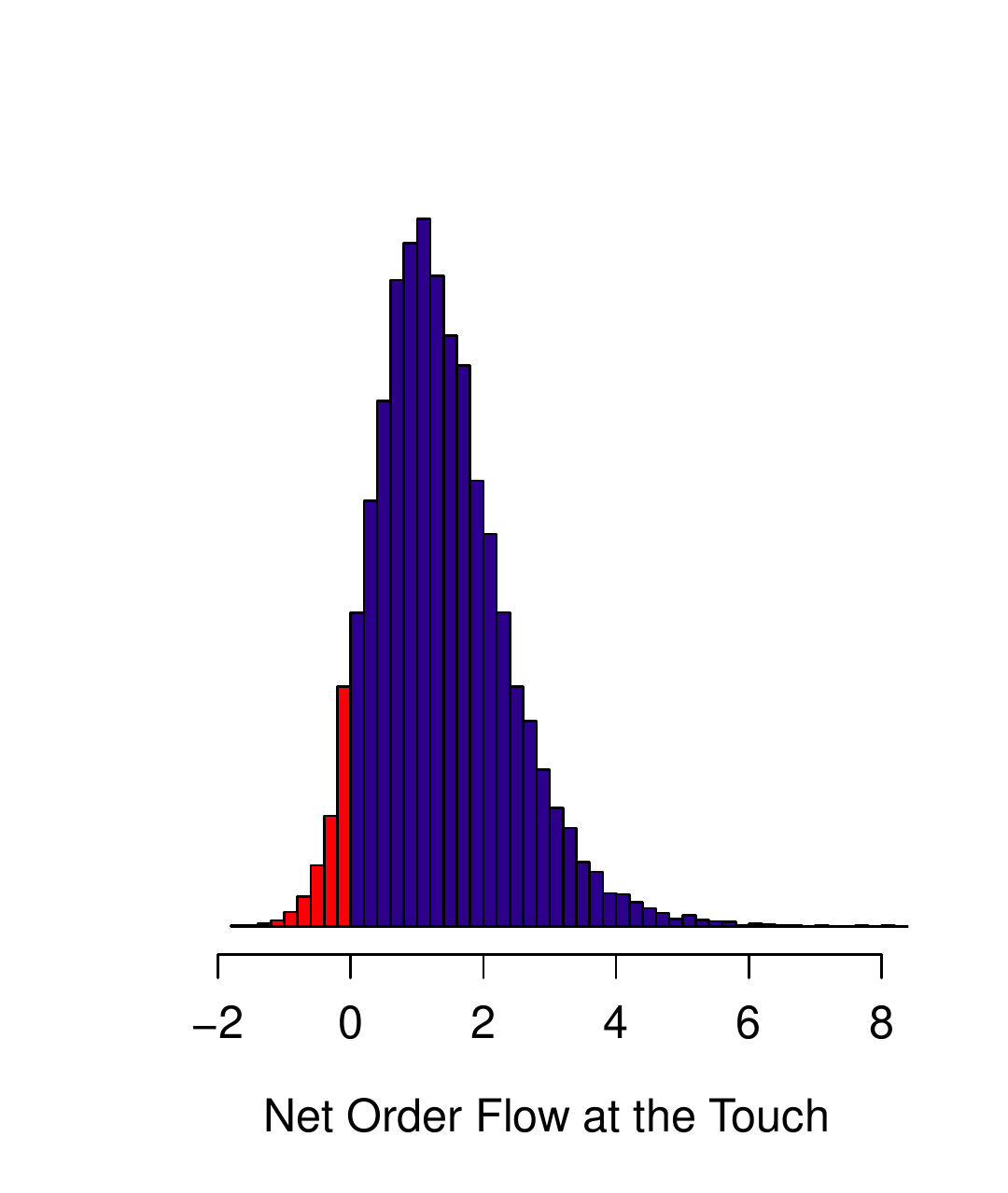} &
			\includegraphics[height=2in,trim=0in 0.2in 0in 0.3in]{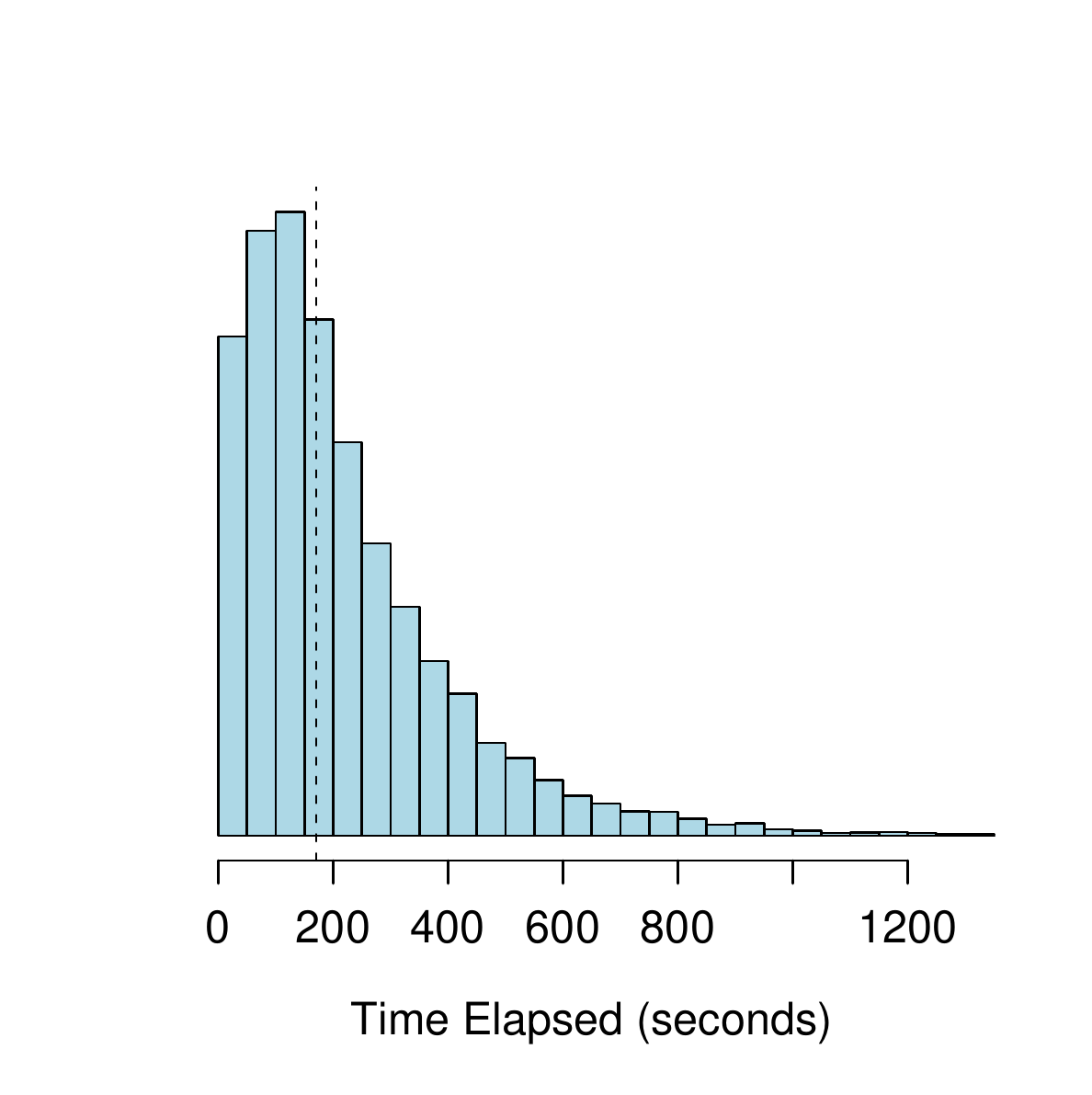} &
            \includegraphics[height=2in,trim=0in 0.2in 0in 0.3in]{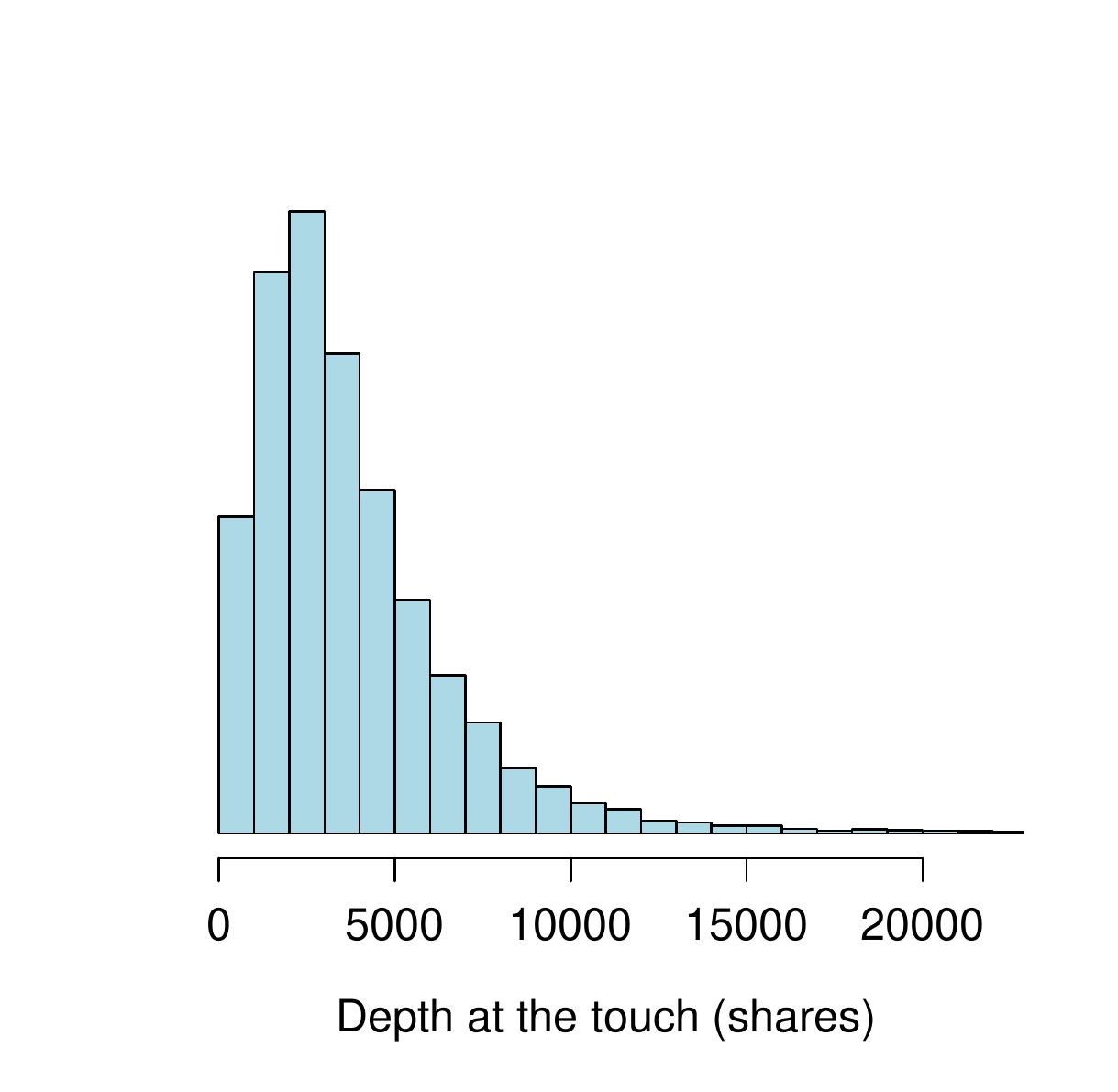}
		\end{tabular}
		\caption{\emph{Left:} Volume-normalized limit order flow at the touch $VL_k/V$ (net of additions and cancellations). About 5.5\% of buckets had negative $VL$ indicated in red. \emph{Middle:} Time elapsed during volume buckets $\Delta \tau_k$; the mean duration is indicated by the dashed line. \emph{Right:} one-sided depth $D^j_1$ at the touch (combined Bid/Ask information). Figures drawn for ORCL over $100$ trading days with buckets of $V=20K$ traded volume (a total of 9229 buckets).
			\label{fig:dat1.5}}
	\end{figure}

To illustrate how volume aggregation works, we describe it in the case of ORCL. We use $V=20000$ which is about 1\% of the average daily volume of 1.84M shares traded during 10:00am--3:45pm. In total, we obtain 9229 aggregated buckets over the 100 days of data. Since the average trade size for ORCL is about 800 shares, a typical slice has about 25 trades, and about 500 limit orders. Figure~\ref{fig:dat1.5} shows a histogram of net limit order-flow $VL_k$ (using combined Bid and Ask data) for ORCL. As expected, average limit flow at the touch is typically positive, i.e.~there are more orders entered than cancelled, providing liquidity against executed trades. For ORCL at that aggregation frequency about 5.5\% of the buckets had negative $VL^j$; this number fluctuates in the 2--10\% range across the 6 assets. Also observe that typically the  nominal limit order flow $VL$ is larger than executed volume $V$, hinting at the high rate of ``churn'' in the limit orders, even after netting additions and cancellations.
 The middle panel of Figure~\ref{fig:dat1.5} shows the distribution of the ORCL bucket durations, $\tau_{k+1} - \tau_k$; the median duration is about 170 seconds. Note the strong skew with many buckets that last for 5 or even 10 minutes; the longest bucket covered 29 calendar minutes (during a particularly slow lunchtime on day 19). Notably, the correlation between $VL_k$ and $\tau_{k+1}-\tau_k)$ is statistically $0$, implying minimal relationship between liquidity provided and physical time elapsing within a volume slice. Appendix A provides a summary of the bucketing employed for each asset.

\section{Price Trend and Liquidity State}\label{sec:results}
As discussed, at the meso-scale the most important quantities are the order flows, i.e.~$VM$ and $VL$ from  \eqref{VMeq}-\eqref{VLeq}. In this section we document their relationship to the fundamental financial output of the LOB: the mid-price change $\Delta P_k$.

	\begin{figure}[ht]
		\centering
		\begin{tabular}{cc}
            \includegraphics[height=2.4in,trim=0in 0.2in 0in 0.3in]{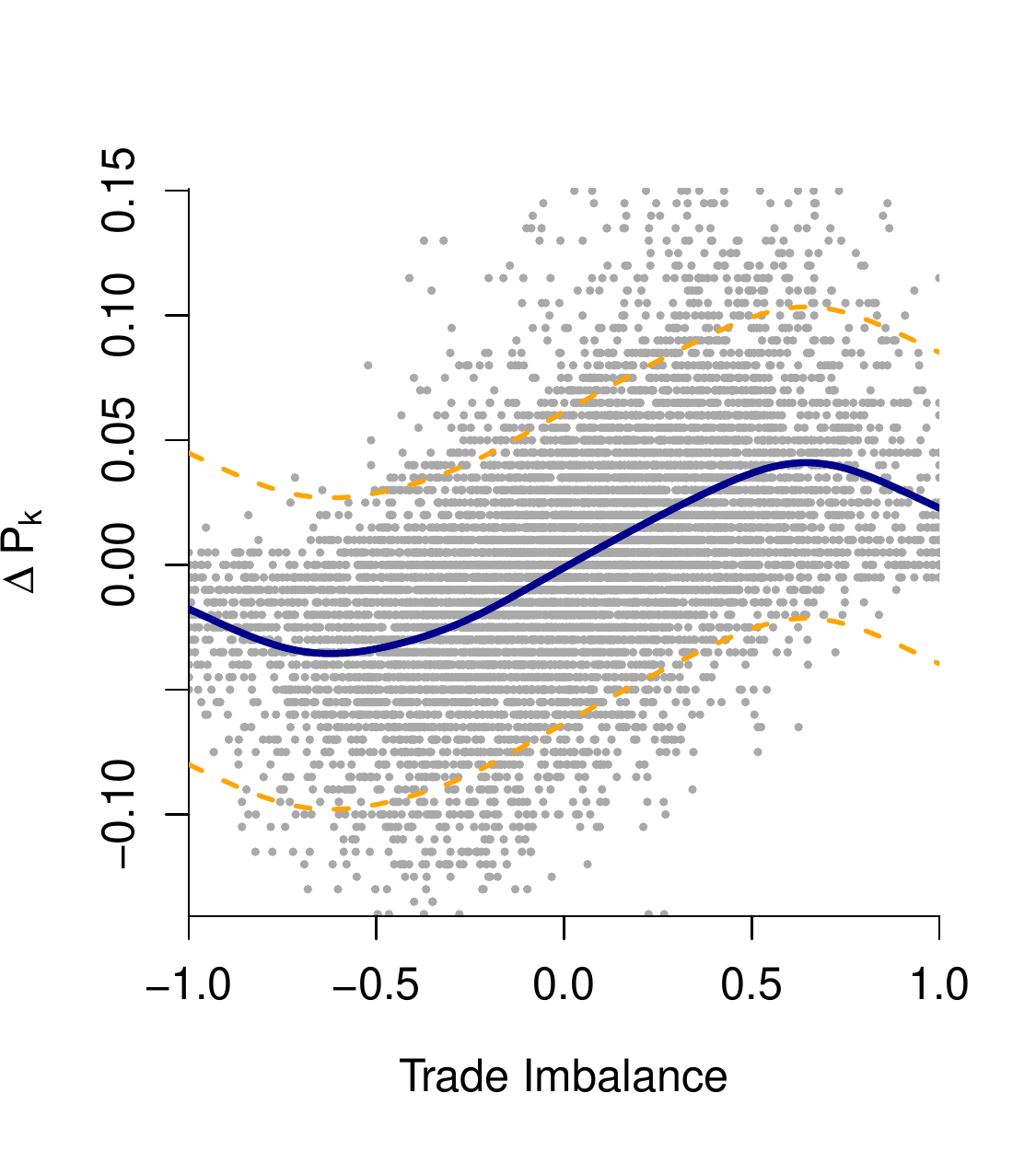}  & 
        	\includegraphics[height=2.4in,trim=0in 0.2in 0in 0.3in]{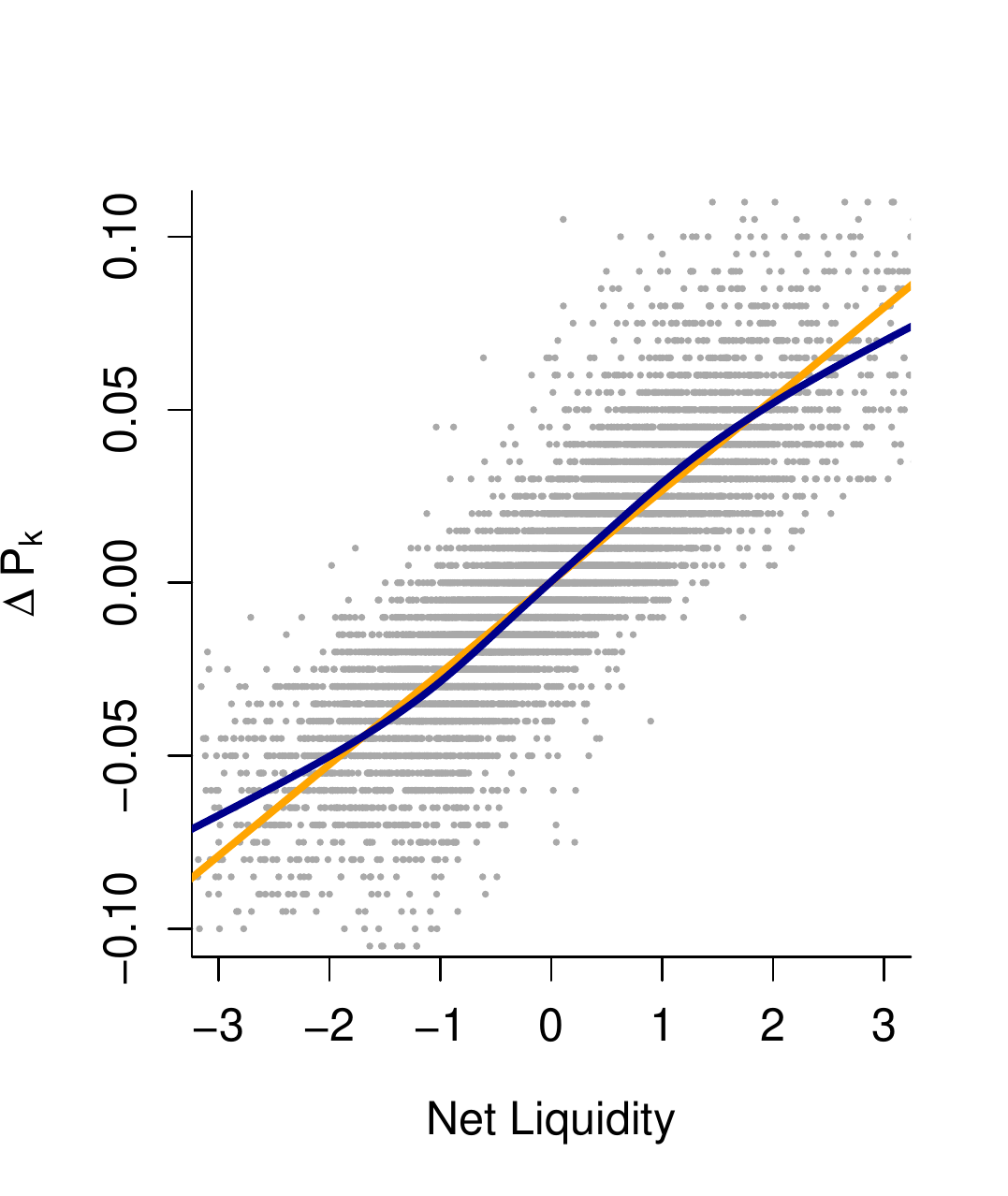}
		\end{tabular}
		\caption{\emph{Left}: Price change $\Delta P_k$ plotted against trade imbalance $TI_k$  for TEVA for the first 100 days of 2013 (9954 buckets with $V=12000$). The fitted curve is based on \eqref{nonlineq}. \emph{Right:} Price change against net liquidity provision for NTAP over buckets of $V=8000$. We plot the linear fit from \eqref{eq:lin-VL} (light line) and the non-parametric GAM model $g(\cdot)$ based on $\Delta P = g(NetLiq) + \eps$, with $NetLiq$ from \eqref{eq:netliq} (dark curve). Since $\Delta P_k \in \{0.005 k : k \in \mathbb{Z}\}$, the $y$-axes are discretized by half-ticks.
			\label{fig:SvsTI}}
	\end{figure}

Based on the price formation mechanism originating in the LOB, the changes in the mid-price are driven by the respective liquidity consumption and provision. A simple narrative ties $\Delta P_k$ to the market demand for the asset, exemplified by the trade imbalance $TI_k$. Self-evidently, if there are more market buys than sells, the price is expected to rise, and if $TI_k < 0$ we expect $\Delta P_k < 0$. A basic model assumes a \emph{linear} relationship between $\Delta P$ and $TI$. A theoretical justification for this is provided in the Obizhaeva and Wang~\cite{ObizhaevaWang} framework and uses a combination of permanent and transient price impact to link executed volume to price move. The basic premise is that limit orders refill the book over time, but take time to be entered. Therefore, price changes are driven by ``too many too fast'' trade executions, and quantified by the LOB resilience (e.g.~the rate of the respective exponential recovery), which yields the sensitivity of $\Delta P$ to $TI$. 

The left panel of Figure~\ref{fig:SvsTI} shows a scatterplot of $\Delta P$ against $TI$ over buckets of 1\% ADV (12,000 shares) for TEVA. While a positive relationship can be seen as expected, the link is clearly \emph{non-linear}. In fact, there is an S-shape behavior, whereby for strongly one-sided markets, $TI$ close to $\pm 1$, the expected price change becomes constant or even declines. This puzzling result in fact indicates intelligent order-placing: traders take into account the state of the LOB when executing orders and hence induce dependency between order flow and price moves. In other words, 12,000 shares executed consecutively in the same direction is an infrequent occurrence for TEVA, and is usually precipitated by a deep/resilient book at that moment. Another important feature of Figure~\ref{fig:SvsTI} is the dispersion in the observed price change $\Delta P_k$ conditional on concurrent trade imbalance $TI_k$. For example, conditional on $TI_k=0.5$, we observe that $\Delta P_k | TI_k=0.5$ ranges from $-0.08$ to $0.16$ for buckets across our sample period, which is financially material. This indicates that on its own trade imbalance has low predictive power, i.e.~price formation is driven by much more than the net executed volume.

To statistically capture the relationship between price change over each volume slice $\Delta P_k= P_{k+1}- P_k$ and the concurrent trade imbalance $TI_k$ we fit a nonparametric model of the form
	\begin{equation}\label{nonlineq}
	\Delta P=g(TI)+\epsilon,
	\end{equation}
	where $g$ is the link function to be estimated and $\epsilon_k$ are i.i.d.~regression residuals. Specifically to account for the observed non-linear dependence we use a generalized additive model (GAM), taking $g(\cdot)$ in \eqref{nonlineq} to be a penalized regression spline that is fit via cross-validation. The analysis is done using the \texttt{R} package \texttt{mgcv} \cite{gamPackage} and is carried out for each stock on volume buckets from the entire sample period, excluding the rare buckets where a single large market order filled the entire bucket (leading to a bucket with zero duration). An example of the resulting fit $TI \mapsto g(TI)$ is shown in the left panel of Figure~\ref{fig:SvsTI} illustrating the aforementioned S-shape.

\begin{figure}[ht]
		\centering
		\begin{tabular}{lr}
            \includegraphics[height=2.4in,trim=0in 0.2in 0in 0.3in]{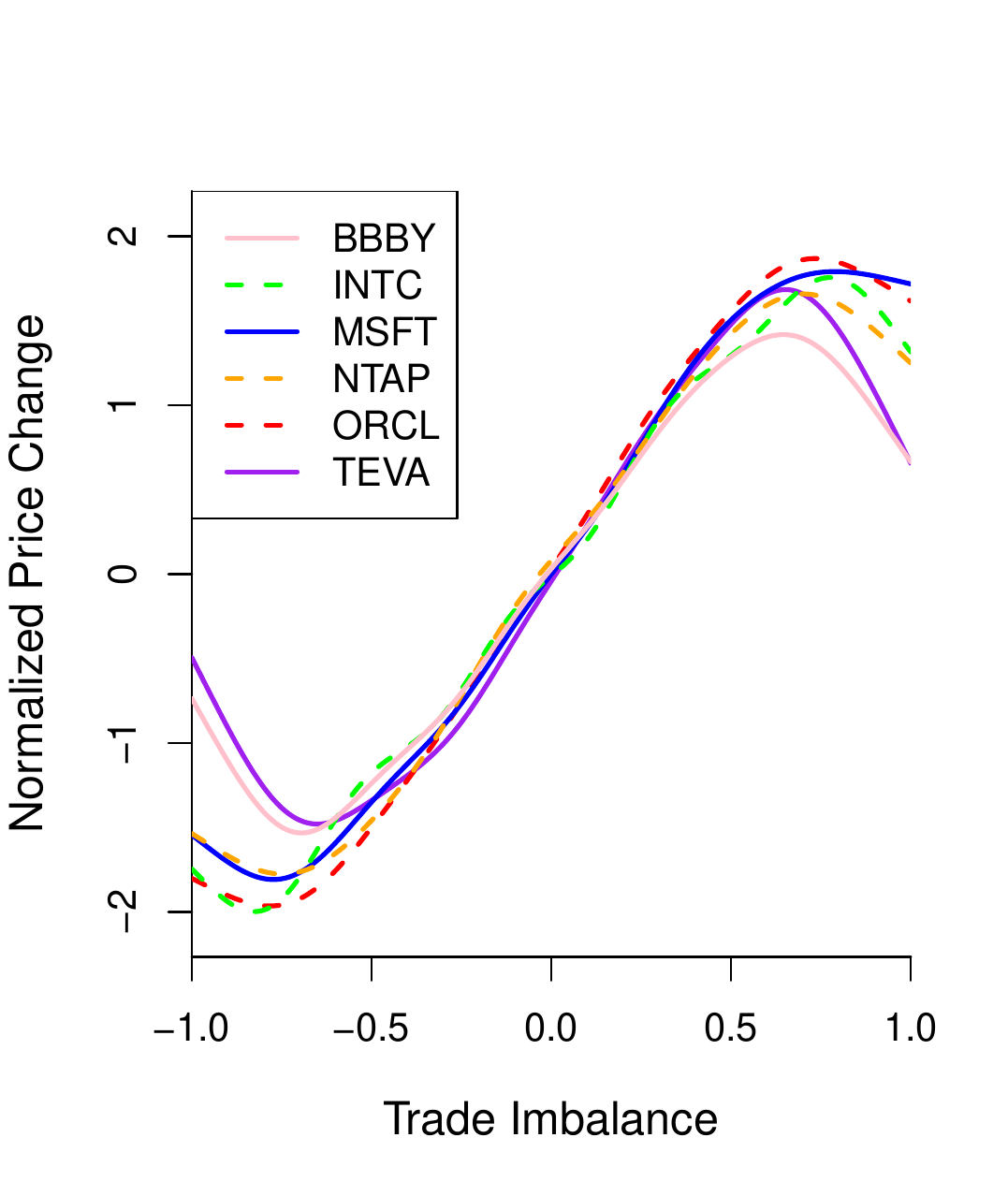} &
			\includegraphics[height=2.4in,trim=0in 0.1in 0in 0.3in]{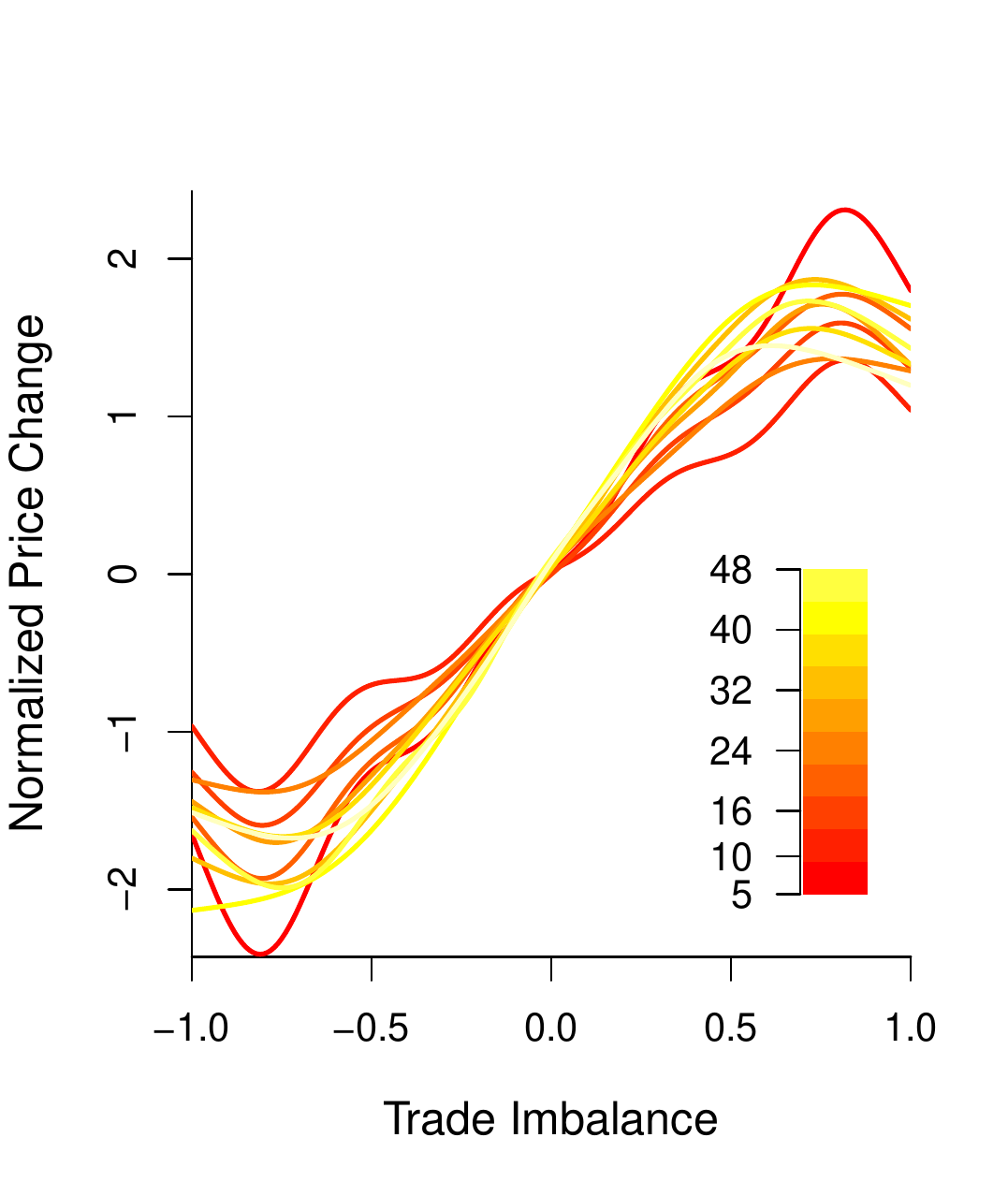}
		\end{tabular}
		\caption{Relationship between normalized price change $\Delta P_k/Ave( |\Delta P_k|)$ and bucket trade imbalance $TI_k$. \emph{Left:} six tickers across the two sample periods; \emph{Right:} for ORCL across 10 different bucket sizes $V$.
			\label{fig:S-curves}}
	\end{figure}

Figure~\ref{fig:S-curves} shows a  cross-section of fitted GAM models~\eqref{nonlineq} across assets (left panel) and bucket sizes $V$ (right panel). We find that the S-shape observed in Figure~\ref{fig:SvsTI} is highly persistent, so that it is a \emph{fundamental stylized feature} of the liquid, large-tick LOBs under consideration. In order to compare the fitted $g$'s, we normalize the $y$-axis measuring price changes by the respective mean absolute price move. This yields a consistent way of mapping trade imbalance, which is by definition in the range of $\{-1,1\}$ to the relative price move $\Delta P_k/Ave(|\Delta P_k|)$. 

The above findings are in contrast to the widely used linear price impact framework that can be traced back to \cite{ObizhaevaWang} and has since appeared in numerous optimal execution/market-making models, e.g.~\cite{alfonsi2010optimal, gatheral2012transient, alfonsi2012order}. Part of the discrepancy can be attributed to our volume-based bucketing which through normalizing $TI$ to be in $[-1,1]$ crystalizes the underlying nonlinearity. From a different direction, our finding is consistent with the documented  volume-concave impact of \emph{individual} market trades which implies a concave relationship between $|TI|$ and $\Delta P$~\cite{eisler2012price}.

\begin{remark}
    One approach to improve the goodness-of-fit of \eqref{nonlineq} is to fit a separate regression model for each trading day. Indeed, it is often assumed that the LOB behavior is driven by a daily cycle. We found that while  daily models do reveal significant day-to-day variation in the shape of the best fit curve (cf.~\cite{cartea2015incorporating}), the resulting $R^2$ is only marginally better compared to fitting a single model over all days. For example, the $R^2$ based on individual day fits for MSFT 1\% ADV improves from 46.9\% in Table \ref{tbl:rsq} to $53.3\%$. So the significant noise in \eqref{nonlineq} is not simply the result of inter-day variation, but rather shows that liquidity provision is stochastic on an intra-day basis. Furthermore, checking the residuals $\widehat{\epsilon_k}$'s from \eqref{nonlineq} fitted to the whole sample period, we find that they are reasonably Gaussian and exhibit minimal temporal autocorrelation.
\end{remark}

\subsection{Limit Flow}

Fundamentally, price is driven by the \emph{net} liquidity supply; looking just at traded volume as was done in \eqref{nonlineq} only considers liquidity consumption. Thus, a natural next step is to look at both market orders that consume liquidity, and limit orders that provide liquidity. For the latter, we concentrate  on orders added/cancelled at the present bid/ask touch which are by far the most relevant given the liquid LOBs under consideration. To capture this idea, we fit the linear model
\begin{align} \label{eq:lin-VL}
  \Delta P = \alpha_0 + \alpha_1 TI + \alpha_2 (VL^B - VL^A) + \eps,
\end{align}
which connects price change to net executed volume and net limit order flow. The main finding is that \eqref{eq:lin-VL} yields an \emph{excellent} fit. Moreover, the linearity in \eqref{eq:lin-VL} is \emph{intrinsic}, i.e.~taking into account the limit flows $VL$ completely removes  the previously documented $S$-shape. Let us define
\begin{align}\label{eq:netliq}
  NetLiq := TI + \beta \cdot VL,
\end{align}
where $\beta = {\alpha_2}/{\alpha_1}$ is the ratio between the impact of a market order vis-a-vis that of a touch limit order. The right panel of Figure ~ref{fig:SvsTI} shows (for a different ticker, to showcase the generality of the phenomenon)  that $NetLiq$ captures the fundamental price formation story of liquidity provision, yielding $\Delta P \propto \Delta NetLiq$ (the estimated intercept coefficient $\alpha_0$ in \eqref{eq:lin-VL} is statistically indistinguishable from zero). In the Figure we also compare the linear and GAM-based fits of $\Delta P$ against $NetLiq$  which are found to be nearly identical, confirming the linearity postulated in \eqref{eq:lin-VL}).

 \begin{table}
$$  \begin{array}{c|ccc|ccc} \hline
 & \multicolumn{3}{ c| }{ \text{Only TI } \eqref{nonlineq}} & \multicolumn{3}{ c }{ \text{Net Liquidity } \eqref{eq:netliq}} \\
 \text{Ticker} & 0.25\% \text{ADV} &  1\% \text{ADV} & 2\% \text{ADV} & 0.25\% \text{ADV} &  1\% \text{ADV} & 2\% \text{ADV}\\ \hline\hline
  \text{BBBY}  &  0.290 & 0.270 & 0.243 & 0.493 & 0.596 &  0.590 \\
  \text{INTC}  & 0.303 & 0.403 & 0.421 & 0.371 & 0.682 & 0.767 \\  
   \text{MSFT}  & 0.370 & 0.469 & 0.485 & 0.579 & 0.785 & 0.765 \\
  \text{NTAP}  & 0.293 & 0.315 &0.302 & 0.470 & 0.712 & 0.751 \\
 \text{ORCL}  & 0.320 & 0.314 & 0.276 &  0.449 & 0.708 & 0.750 \\
   \text{TEVA}  & 0.269 & 0.276 & 0.269 & 0.520 & 0.607 &  0.592 \\
  \hline 
\end{array}
$$
\caption{$R^2$ scores from the nonparametric fit against $TI$ in \eqref{nonlineq} compared to $R^2$ scores from a linear fit against $TI$ and $VL$ in \eqref{eq:lin-VL}. \label{tbl:rsq} }
\end{table}

Incorporating order flows not only removes the nonlinear price response, but also dramatically improves the goodness-of-fit, see Figure~\ref{fig:SvsTI} again. Table \ref{tbl:rsq} compares the $R^2$ scores from \eqref{nonlineq} and \eqref{eq:lin-VL}. We observe a major improvement from about $R^2 \simeq 0.4$ when using only $TI$ as a predictor for $\Delta P$, to $R^2 \simeq 0.7$ when using both $TI$ and $VL^j$. This suggests that the predictive power of $VL^j$ is \emph{at least as significant} as of the executed volume. Table \ref{tbl:rsq} also shows that goodness-of-fit improves as we consider larger bucket sizes, implying that the other predictors get averaged out over time (i.e.~are primarily about the microstructural effects of the LOB).

While mechanically net liquidity is simply
$
VM^A - VM^B + VL^B - VL^A,
$
it is commonly assumed that the price impact of limit orders should be less than that of market orders. This is precisely the reason for using separate coefficients $\alpha_1, \alpha_2$ in \eqref{eq:lin-VL}. The resulting ratio $\beta$ that is used to construct $NetLiq$ in \eqref{eq:netliq} is another \emph{fundamental stylized feature} of the meso-scale.
Table \ref{tbl:coef} reports the estimated $\hat{\beta} := \widehat{\alpha_2}/\widehat{\alpha_1}$ across assets and bucket sizes, which can be observed to be highly stable across all cases. As expected, the respective price impacts are highly non-equal, namely the relative effect of a limit order is only about 50\%--70\% of a market order. Again, we observe that $\beta$ is larger for larger $V$'s: on the macro-scale price formation is mechanically determined by total liquidity supplied/consumed; on the meso-scale there are further effects coming from LOB shape and shorter-term patterns.

\begin{remark}
We also tested for the relative price impact of limit order additions and cancellations by further decomposing into  $VL^j = VL^{j,+}-VL^{j,-}$. Statistically, the price impact of the two latter order types was the same, giving minimal improvement to the fit in~\eqref{eq:lin-VL}. However, as we will show below, cancellations are important for capturing periods of scarce liquidity.
\end{remark}


 \begin{table}
$$  \begin{array}{c|ccc} \hline
 \text{Ticker} & 0.25\% ADV & 1\% ADV & 2\% ADV\\ \hline \hline
  \text{BBBY}  &  0.39 & 0.55 & 0.58 \\
  \text{INTC}  & 0.20 & 0.42 & 0.55 \\
  \text{MSFT}  &  0.41 & 0.67 & 0.59 \\
  \text{NTAP}  &  0.33 & 0.64 & 0.68 \\
  \text{ORCL}  & 0.24 & 0.57 & 0.68 \\
    \text{TEVA}  & 0.49 & 0.66 & 0.66 \\ \hline
\end{array}
$$
\caption{Price impact $\beta$ of touch Limit Orders relative to Market Orders based on  \eqref{eq:netliq}. \label{tbl:coef}}
\end{table}

 The idea of \eqref{eq:netliq} echoes the analysis in Cont et al.~\cite{cont2012price}, who however worked with time-based buckets (of 10 seconds, so generally shorter than ours) and combined all flows into a single predictor, i.e.~pre-assigned equal price impact $\alpha_2 = \alpha_1$ or $\beta=1$ to market and limit orders. Our analysis suggests that $\beta$ is significantly less than 100\%, which seems to be closer to market practice. On the one hand, intuition suggest that executed trades carry more information, i.e.~are more relevant for questions of adverse selection or informed trading and hence contribute more to price formation. On the other hand, the nominal volume of limit orders  $VL$ is 1--3 times larger than nominal volume of $VM$; without separate coefficients, $\Delta P$ would be unduly affected by limit orders. The above findings are also consistent with the event-by-event analysis done in \cite{eisler2012price,HautschHuang12} who found that the impact of market trades (measured by the temporal impulse response function) for large-tick Nasdaq stocks is 2--4 times larger than that of limit orders. 

\begin{remark}
 We also mention the less-known work of \cite{Hopman07} who proposed a different way of defining Net Liquidity by using power-weighing, namely considering $\sum_{i: \tau_k < T^L_i \le \tau_{k+1}} (O^L_i)^{\alpha} 1_{ \{S^L_i = p^j_1( T^L_i)\}}$ in \eqref{VLeq}, with the power coefficient $\alpha \in [0,1]$. In particular, he advocated looking at square-root-volume ($\alpha=0.5$) or just order counts ($\alpha=0$) when constructing the $NetLiq$ measure. On the latter point, \cite{benzaquen2016dissecting} used the \emph{number} of market trades to estimate permanent price impact.
\end{remark}

\subsection{Limit Flows vs Market Flows}\label{sec:vl-vm}
The previous subsection documented that price change is driven by a weighted average of $VM$ and $VL$, with the link being essentially linear. This motivates the analysis of the relationship between trade imbalance and limit order flows, shown for a representative ticker in Figure~\ref{fig:VLvsTI} below. We separately plot $VL^A$ and $VL^B$ against TI to highlight the inherent \emph{nonlinear} dependence between the two sides of the book. Indeed, when the market is unbalanced (large $|TI|$), the book behaves asymmetrically: the net limit order activity is stable on the passive side, while liquidity provision on the active side declines. We conjecture that this effect is due to market-makers competing to avoid adverse selection, leading to diminished new limit orders and existing order cancellations. Thus, Figure~\ref{fig:VLvsTI} shows that there is a hockey-stick-response curve $VL^j = g^j(TI)$ on each side of the book. For example, looking at $VL^A$, there is a ``base-line''  level of liquidity provision that takes place as long as $TI < 0.3$ or so (i.e.~less than 2/3 of execution are buys). However, when the market is dominated by Buys, $VL^A$ tends to be lower, i.e.~the Ask side is replenished less than usual under  upside price pressure. For some stocks (BBBY, TEVA see Figure~\ref{fig:allVLvsTI} in Appendix B) we see an upside-down V-shape, whereby $VL^A$ also falls for $TI$ close to $-1$, i.e.~the Ask side ``hibernates'' under intense sell pressure. Mirror effects take place on the Bid side.  These findings suggest  that LOB resilience (and hence price impact) is essentially driven by \emph{one-sided flows} rather than aggregate two-sided metrics. For example, when net limit order flow at the best bid $VL^B$ is large, the price is unlikely to move lower even if trade imbalance is negative.

Figure~\ref{fig:VLvsTI} also shows that large upside price moves tend to coincide with (i) positive $TI$; (ii) smaller than usual $VL^A$ and (iii) larger than usual $VL^B$; the converse holding for downside price moves. Hence when prices move rapidly, both the market and limit order flows are highly unbalanced, lending credence to weak resilience manifesting itself as a  ``wrong-way'' correlation between $VM$ and $VL$. Given the very large number of heterogeneous players participating in trading activity, it is difficult to draw any conclusions as to the respective causality. Whether HFT market-makers are predicting one-sided market flow or reacting to it, liquidity provision via $VL$ seems to reflect liquidity consumption in $VM$.

	\begin{figure}[ht]
		\centering
		\includegraphics[height=2.1in, trim=0.1in 0.1in 0.1in 0.2in]{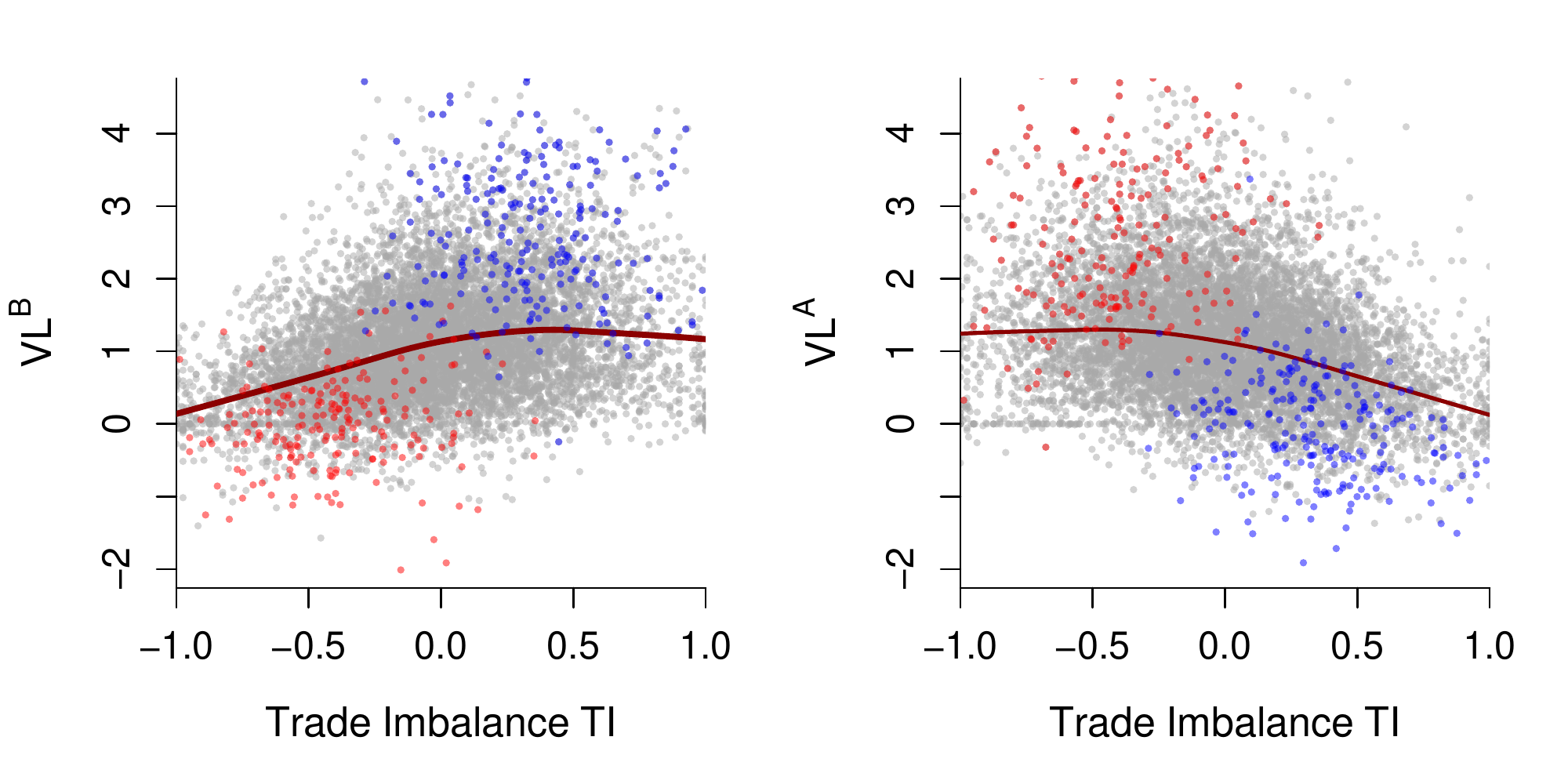} 
		\caption{Scatterplot of the net limit order flow at the best bid $VL^B_k$ (left) and best ask $VL^A_k$ (right) against the trade imbalance $TI_k$ across the 8886 buckets of $V=8000$ for NTAP. The $y$-axis is normalized via $VL^j/V$. Red (blue) points indicate volume buckets with price decrease (increase) $\Delta P_k$ of more than $0.07$. The solid lines indicate the average response $g^j(TI)$ obtained from a spline-based GAM fit to $VL^j = g^j(TI) + \eps$. \label{fig:VLvsTI}}
	\end{figure}

\begin{remark}
An extreme example of weak resilience was documented in \cite{lehalle2012sawtooth} in the context of the market events of July 12, 2012. On that day four large-cap US stocks exhibited an unusual trading pattern: heavy buying following by heavy selling in a predictable fashion over $30$ minute intervals. The result was a sawtooth pattern in the asset prices, apparently generated by significant price slippage for the aggressive buyers/sellers.  Lehalle et al.~\cite{lehalle2012sawtooth} conclude that this weak LOB resilience was driven by liquidity providers who anticipated one-sided market order flow and were jockeying for queue priority via rapid cancellation of their resting top-level orders after each execution.
\end{remark}

		\section{Liquidity Predictors}\label{sec:liquidity}
The results in the previous section strongly indicate that including concurrent limit order flow $VL_k$ along with trade imbalance $TI_k$ dramatically improves the goodness-of-fit for predicting $\Delta P_k$. Nevertheless, for purposes of optimal execution the role of trade imbalance remains paramount.
 Indeed, the execution algorithm displaces some of the market orders, so that the order scheduler in effect partially controls the trade imbalance $TI_{k+1}$ in the upcoming bucket. Hence, forecasting the conditional price change $\Delta P_{k+1} | TI_{k+1}$ is crucial in  scheduling the next slice.

Unlike the time slicing approach, each bucket in our setup contains an identical amount of traded volume. Thus the difference between observed price change $\Delta P$ for two buckets with similar trade imbalance $TI$ must be due to differences in LOB resilience and/or shape. Therefore, the findings in the previous section
lead to the natural decomposition of price formation into two components: (1) \textit{price trend}, which is primarily about the supply and demand for the asset captured in trade imbalance $TI_k$, and (2) \textit{liquidity}, which is seen in the deviation of the observed $\Delta P_k$ from the best fit curve between the two variables, and has been already shown to strongly depend on $VL^j_k$. In this section we seek to more systematically identify the key variables important for explaining LOB liquidity on the meso-scale. This is done by constructing a series of regression models taking in a variety of potential covariates, and quantifying their  significance and effect.
The following Section \ref{sec:scarce} then concentrates on
predicting the periods of scarce liquidity that lead to out-sized price impact.

For the above purpose, we utilized four classes of statistical models: (i) classical linear models (LM) using stepwise selection criteria; (ii) LASSO models fit via cross-validation; (iii) MARS models; (iv) random forests (RF). Generically, we model
\begin{align}\label{reg2}
\Delta P = \sum_r \phi_r( \mathbf{X})+\epsilon,
\end{align}
where $\mathbf{X}$ is the vector of all the covariates and the basis functions $\phi_r$ are adaptively picked by the model. We focus our attention on the primary effect of various predictors and therefore take $\phi_r$ to be low-dimensional in $\mathbf{X}$: they are linear for LM and LASSO, of the hockey-stick form $(X^{(r)} - K_r)_+$ for MARS, and indicator functions of hypercubes for RF. Thus, except for RF no interactions among covariates are a priori considered.

 The three model frameworks beyond LM offer complementary perspectives on variable importance and ultimately model selection. LASSO uses $L^1$-regularization to drop less relevant predictors giving a crisp quantification of variable importance and easily interpretable coefficients for covariate impact. Multivariate adaptive regression splines (MARS)  aims to identify nonlinear relationships through using piecewise linear hinge basis functions. It can also search for statistically significant predictors from low-degree interactions among covariates. Finally, RF constructs a black-box nonlinear model by randomly considering subsets of predictors to obtain a piecewise constant relationship. Random Forests yield only indirect measures of variable importance or contribution via model averaging, but makes minimal parametric assumptions about the functional form of covariate impact.
Note that our goal is \emph{not} to maximize predictive performance, but to obtain interpretable qualitative analysis. Hence, we do not explore fully non-parametric ``machine learning'' tools.

To build a statistical description for $\Delta P$ we utilized  both static measures based on the LOB state at bucket times $\tau_k$, as well as dynamic quantities measured over time.
The static predictors tested in our regression models included the measures of LOB depth and shape described in Section \ref{sec:Liq}, specifically price impact $PI_N$, impact slope $S$, cumulative depth $D_i$ for $i=1,2$ and book imbalance $BI$. For completeness, we also included time-of-day $\tau_k$. All static quantities were measured at the start of the volume slice i.e.~the price change $P_{k}-P_{k-1}$ from the activity during $[\tau_{k-1}, \tau_{k}]$  was regressed against static variables measured at time $\tau_{k-1}$.  Dynamic quantities included the contemporaneous net limit flows $VL^A_k, VL^B_k$, as well as lagged quantities: trailing limit flow $VL^j_{k-\ell}$, lagged price change $\Delta P_{k-\ell}$, and lagged trade imbalance $TI_{k-\ell}$ over the previous $\ell$ volume buckets, where $\ell \in 1,5,10,20$. Also included was the exponentially weighted moving average of recent trade imbalance, first introduced in \cite{bechler2014optimal},
	\begin{align}\label{eq:ewma}
	TIMA^{(\beta)}_{i+1} := e^{-\beta|O^M_i|} TIMA^{(\beta)}_i +(1-e^{-\beta|O^M_i|})\sgn(O_i^M),
	\end{align}
	where $i$ indexes arrival of market orders of size $|O^M_i|$ and $\beta$ dictates the exponential weights; we set $\beta = 0.5/V$. 

\textbf{Important Covariates. }
Our results indicate that the trade imbalance $TI$ and the limit order flows $VL^j$ are by far the most statistically significant predictors for $\Delta P$.  Further  predictors that were consistently picked by the models include the price impact $PI^j$, the depth at the top-2 levels $D_2$, and the proportion of limit order cancellations $PC^j := VL^{j,-}/VL^j$. For the latter, we observe that when cancellations at the bid dominate additions, $PC^B > 0.5$, the price can move lower on very little volume. Weaker predictors (picked by some models but not all) are Book Imbalance $BI$ and Impact Slope $S$. There are additional covariates that were  statistically significant for one or two of the tickers, but these yield negligible improvement to overall model fit. See Table~\ref{tbl:rf-predictors} in Appendix~B for a summary of variable importance.

The RF models also assigned strong statistical significance to the interactions $TI \times PI$, $VL^j \times PI^j$ (one-sided interaction between limit flow and book impact) and $TI \times S$. Since both $PI$ and $S$ are in units of ticks/share, the latter two terms are denominated in ticks, and provide an impact-normalized measure of market trend. Indeed, the static LOB metrics capture the shape of the book and hence \emph{modulate} the effect of the major predictors in the flows. The relevance of the above cross-terms was confirmed by a LASSO model where we manually added all possible interactions between $TI, VL^A, VL^B$ and the impact covariates $PI$ and $S$. As an additional check, no further interactions  (considered by building a degree-2 MARS model) were deemed significant. Overall, using the $R^2$ as a simple goodness-of-fit metric, we find that it starts out at $R^2 \simeq 30\%$ when regressing $\Delta P$ against just $TI$; rises to about $R^2 \simeq 70\%$ once $VL^A, VL^B$ are included; and reaches about $R^2 \simeq 80\%$ after adding in the mentioned LOB shape metrics. Another 1-2\% gain in $R^2$ is possible by including all the covariates listed, including their lagged versions.

Of note, top level depth $D_1$ was statistically insignificant in the regression tests. Given the amount of activity in each volume bucket it is not surprising that ``shallow'' LOB measures capturing only the top of the LOB are not all that useful. Instead, the models identify (partly due to the underlying collinearities) a mix of the deeper LOB shape metrics, including the top-2 levels volume $D_2$, the price impact $PI_N$ (which looks at 3-5 levels deep, $N \simeq Ave(D_4)$), and the impact slope $S$ (a cost-averaged depth over $\simeq 4$ levels). We note that for more liquid tickers  (such as MSFT), $D_2$ is more relevant, while for less liquid ones the models emphasize $PI$.
 Another interesting find was the (moderate) significance of trailing moving average of executions $TIMA$ from \eqref{eq:ewma}, with a consistently negative relationship to $\Delta P_k$. The inclusion of $TIMA$ appears to be capturing the tendency of stock prices to, more often than not, retrace recent movement. More precisely, when $TI$ leans with the prevailing trend ($TIMA$) the impact on price is less.

 Returning to a more qualitative description, Figure~\ref{fig:pairs} (Appendix~C) shows paired scatterplots across 9229 buckets of ORCL for limit flows $VL^j$, proportion of cancellations $PC^j$ and impact slopes $S^j$. Some important observations are:
\begin{itemize}[leftmargin=.1in]
  \item Strong negative correlation $\rho \simeq -0.4$ between $VL^A$ and $VL^B$, meaning that Ask/Bid limit order activity is complementary: at any given moment most of the limit orders go to one side;

    \item Positive correlation ($\rho \simeq 0.45$) between order flow $VL$ and proportion of cancellations $PC$ on the \emph{other} side: as more orders are added, say, on the Ask side, limit orders are also cancelled on the Bid side;

  \item Negative correlation between depth $D$ and order flow  $VL$ on the \emph{opposite side}, so that the less-active side of the book also tends to be more shallow;

  \item Positive correlation between book depth on the two sides, $D^A$ and $D^B$, meaning that the book as a whole fluctuates between being deep and shallow;

  \item Weak positive correlation ($\rho \simeq 0.2$) between cancellation proportions $PC^A$, $PC^B$ on the two sides, dovetailing with the previous point.

\end{itemize}
These complex cross-relationships imply that disentangling the marginal effect of a given predictor is difficult, and a simple additive model, such as MARS or LASSO, should be interpreted with care. For instance, the ask-side depth metric $D_2^A$ is intrinsically correlated with $D_2^B, VL^B, S^A, PC^A$ and so on. Therefore, the counterfactual of changing ask-side depth while keeping all other covariates fixed, captured by a regression coefficient (or partial dependence plot) of $D_2^A$, is not too meaningful.

	\subsection{Scarce Liquidity}\label{sec:scarce}
	
Intuitively, scarce liquidity  describes the situation where large price moves occur under low volume, indicating weak resilience. Within our framework, the latter  corresponds to an ``abnormally'' large $\Delta P_k$ relative to the contemporaneous trade imbalance $TI$. Pinpointing the confluence of factors that coincide with scarce liquidity is of practical importance due to the potentially large ramifications of  executing during a period of low resilience in the LOB.

 To capture the idea of a ``disproportionately large'' price move relative to executed volume we  remove the effect of trade imbalance to focus on the resulting residuals $\widehat{\epsilon}_k = \Delta P_k - g(TI_k)$ from \eqref{nonlineq}. Namely, we define a scarce liquidity indicator variable $SL^j_k \in \{0,1\}$ in a binary fashion in terms of $\widehat{\epsilon}_k$, so that $SL^j$ occurs on at most one side of the LOB per volume bucket:
	\begin{equation}\label{singular3}
	SL^A_k:= I\{ \widehat{\epsilon}_k\geq \bar{M} \}, \qquad SL^B_k := I\{ \widehat{\epsilon}_k\leq - \bar{M} \},
	\end{equation}
	where we empirically choose the scarcity threshold $\bar{M}$ in terms of the standard deviation of the residuals $\hat{\epsilon}$, specifically $\bar{M}=1.5StDev(\widehat{\epsilon}_{1:K})$. Thus, scarce liquidity occurs on each side of the LOB in approximately 6\% of volume buckets. Our data-driven definition of liquidity is similar to Amihud~\cite{Amihud02}. The pairs plot in Figure~\ref{fig:pairs} in the Appendix color-codes buckets with scarce liquidity, highlighting the relationship of key predictors to $SL^j$.

To quantitatively analyze the occurrence of scarce liquidity we ran a logistic regression model for $SL^A$ and $SL^B$ with the covariates described in the previous section. Compared to the regression models for $\Delta P$ of the previous section, logistic regression is geared towards capturing one-sided LOB effects and also focuses attention on the buckets with more extreme price moves. It can therefore better identify the LOB regimes that cause substantial price moves, compared to \eqref{reg2} that concentrates on fitting the ``typical'' price formation factors. Recall that logistic regression assumes that $SL^j_k \sim Bernoulli(\pi^j_k)$, where the log-odds of $\pi^j_k$ are specified via	\begin{equation}\label{logistic}
	\logit \pi^j= \sum_r \phi^{SL,j}_r(\mathbf{X}).
\end{equation}

This analysis reveals that occurrences of scarce liquidity are driven by a high asymmetry in the limit flows, as well as reduced LOB depth and/or high cancellation rates. For example, the major predictors of $SL^A$ are $VL^B$ (+ve correlation), $TI$ (-), $VL^A$ (-) and $D^{A,B}_2$ (both -). Several of these influences need to come together to explain $SL^A$. Financially, we observe that scarce liquidity tends to occur on a passive and shallow side of the book. For instance, if the Bid side is passive (high $VL^A$ and low $VL^B$), and moreover is less deep than normal (low $D^B_1$), then prices are likely to fall through on the downside, triggering $SL^B=1$. The logistic regression models also show that the
 proportion of limit order cancellations has a high predictive power, so that large price moves are associated with a ``LOB fade''. As in the previous section, the limit order flows were the most statistically significant predictors for  $SL^j$ via \eqref{logistic}. Unlike the previous section, price impact $PI$ had a heightened significance. Specifically, same-side price impact ($PI^j_N$ when testing $\pi^j$) was of approximately equal importance as $VL^j $ for all six tickers. The physical timing $\tau_k$, was also significant, as was bucket duration $\Delta \tau_k$. The fitted models imply that scarce liquidity is more likely in the morning, or when trading is slow (large $\Delta \tau_k$).

Table~\ref{tab4} in Appendix~B reports  goodness-of-fit of the above logistic models using a hold-out test set. The Table shows the relative frequency of $SL^j$ against the predicted log-odds $\widehat{\pi}^j$. We observe that for vast majority of cases the log-odds are low and in those buckets there is indeed almost no cases of abnormally large price moves. On the other hand, when the model predicts scarce liquidity, it is correct in about 70--80\% of the time. We note that due to the low frequency of $SL$, the model predicts less than half of all scarce liquidity occurrences, i.e.~there are more false negatives than true positives.

Table \ref{tbl:rf-predictors} in  Appendix~B lists the significant predictors for the three regressions we ran (for $\Delta P$, $SL^A$, and $SL^B$). In the Table we used the \texttt{importance} function in the \texttt{randomForest} package \cite{randomForestR} to roughly classify given predictors into Most Important (**), Somewhat Important (*) and Unimportant (-). Since the RF model is data-driven and able to capture nonlinear effects the resulting variable importance is more robust compared to other model classes we tried, such as Lasso or MARS which only capture low-degree interactions and have trouble with collinearity. We also manually checked  the predictors selected by LASSO and MARS vis-a-vis those of RF to confirm  covariate importance.

Table~\ref{tbl:rf-predictors} matches the intuition that same-side covariates (especially the impact-dollar metrics $PI^j$ and $S^j$) ought to be more significant for explaining $SL^j$. We also note that the logistic models are more complex (assigning predictive power to more covariates), and moreover  include additional ``two-sided'' predictors, such as bucket duration or price volatility, that would not contribute to $\Delta P$, but make sense for forecasting large price moves.
The observed asymmetric effects imply the need to build separate models for different aspects of the limit-flow/price relationship. The listed significant predictors are remarkably consistent across assets. Moreover, they are complementary/symmmetric on the Ask and Bid side (the respective regressions were run completely independently), indicating that these are \emph{generic stylized facts} for liquid, large-tick LOBs.

\section{Time Series Perspective}\label{sec:ts}
Previous sections carried out a statistical investigation of LOB behavior using the regression paradigm that treats observations as i.i.d.~samples from the ``true population''. Of course, in practice LOB data is sequentially ordered, so time series methods are also appropriate. However, the observed complexity (such as multiple nonlinear relationships) makes constructing multivariate time-series models a major challenge and beyond the scope of this paper. For our purposes we therefore simply present some stylized observations.

\begin{figure}[ht]
		\centering
		\begin{tabular}{cc}
			\includegraphics[height=2.3in,trim=0.2in 0.25in 0in 0.3in]{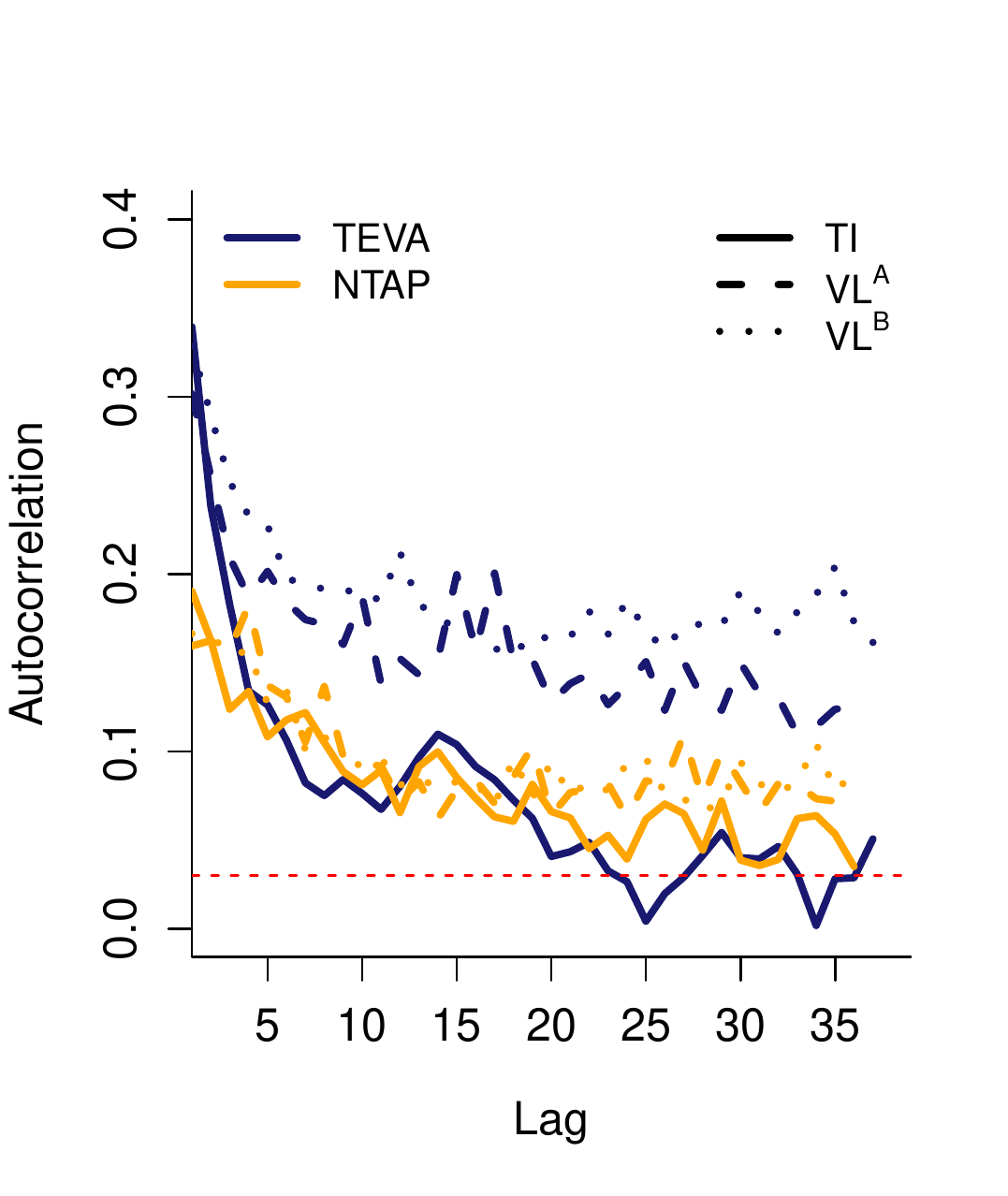} &
			
			\includegraphics[height=2.3in,trim=0in 0.25in 0.2in 0.3in]{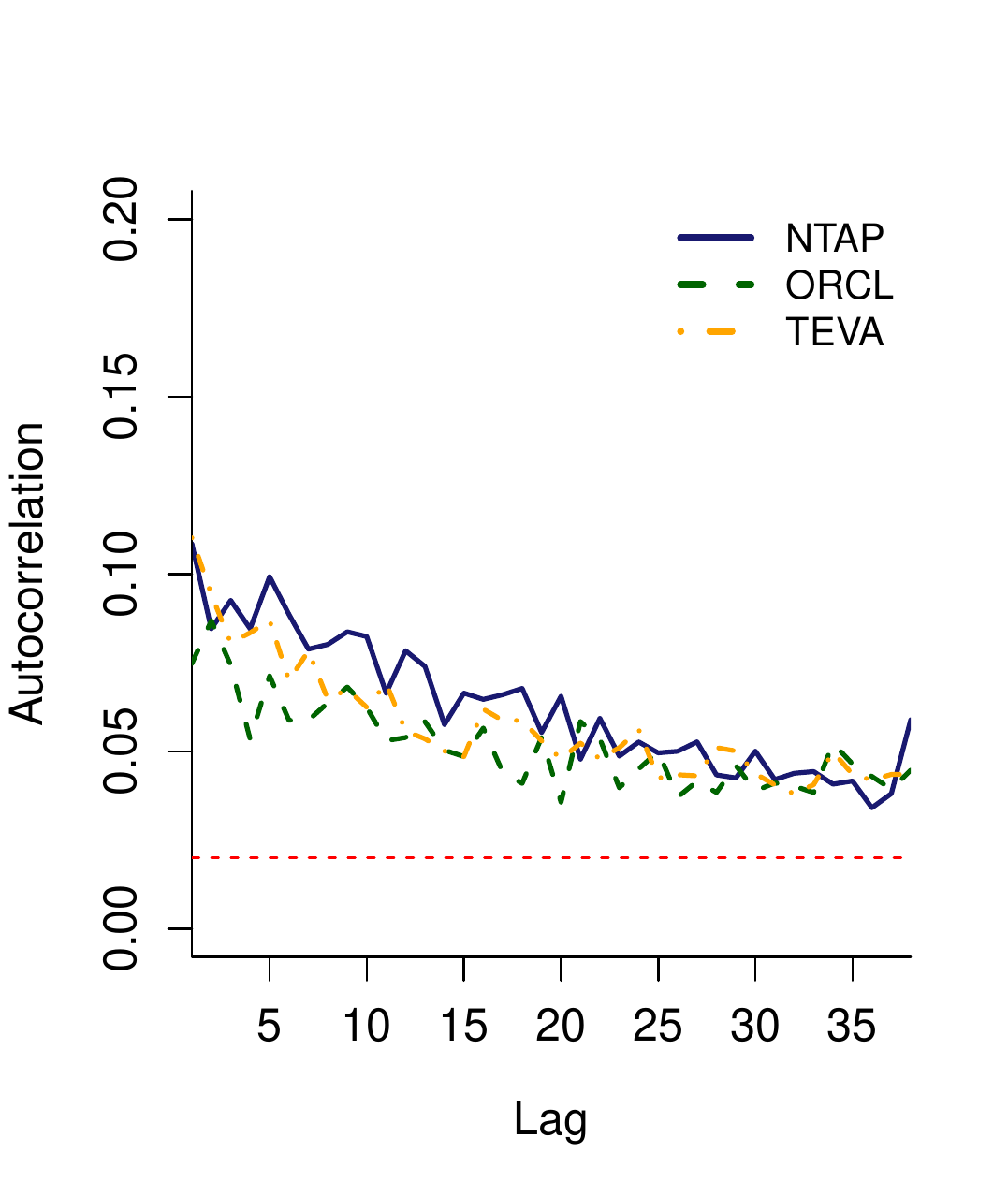} 
		\end{tabular}
		\caption{Marginal autocorrelation functions (ACF) for order flows and scarce liquidity. \emph{Left:} the marginal ACF for the trade imbalance $TI$, as well as one-sided net limit flows at the touch $VL^A$ and $VL^B$. Data is based on 2\% ADV bucketing for NTAP and TEVA. Right: marginal ACF for the scarce liquidity indicator $SL = SL^A + SL^B$. In both panels dashed lines indicate the significance threshold for the presence of autocorrelation. 			\label{fig:acfs}}
	\end{figure}

A basic test is for the presence of auto-correlation which would imply that there are some persistent factors in the observations, and hence serial dependence among buckets that are close in time. We first consider the net order flow variables: $TI, VL^A, VL^B$. We find that there is minimal autocorrelation at the shorter scales (0.25\% ADV buckets), but some persistence is observed on the longer scales and for less-liquid assets. Positive autocorrelation suggests that periods of strong liquidity provision are followed by more of the same, and conversely, low $VL$ tends to persist for multiple buckets.
 For the deeper LOB's, we believe the persistence is still there but a better tool than the ACF diagnostic is needed. Remarkably, the persistence carries through for long lags, indicating presence of long-term memory on the order of hours. Figure~\ref{fig:acfs} illustrates  the non-zero autocorrelation in $VL^j$ for NTAP and TEVA across 2\% ADV buckets. These results echo previous literature (e.g.~\cite{Toth15}) that documents long-term persistence in order flows, possibly even on the scale of days.
 Overall, we note that the computed auto-correlation levels are not too strong (relative to say the very high $R^2$ reported previously), indicating that standard ARIMA models might not fit well.

Turning attention to scarce liquidity, as remarked there is essentially no auto-correlation in the residuals of \eqref{nonlineq} and hence minimal ACF levels in the one-sided $SL^j$ indicators. In other words, the ACF test does not detect any obvious clustering in one-sided scarce liquidity. At the same time, we do find cross-correlation, so that an abnormally large positive price move is likely to be followed by an abnormally large negative $\Delta P_k$. This hints at the ``whiplash'' property of the LOB, where reversions in price change or in $TI$ seem to trigger reduced liquidity provision. The right panel of Figure~\ref{fig:acfs} shows the ACF plot for the combined indicator $SL = SL^A + SL^B$ (volume buckets with scarce liquidity in either direction), which is positively auto-correlated, i.e.~exhibits some temporal clustering.

\subsection{Co-Movement of Ask/Bid Order Flows}
 Recall Figure~\ref{fig:VLvsTI} which presented an aggregated relationship between $VM$ and $VL$. To capture the respective co-movement of flows, we consider the  correlation $\rho^j$  between one-sided limit flows $VL^j$ and $VM^j$, $j=A,B$. Lower levels of $\rho^j$ are indicative of increased cancellations and hence less liquidity, while positive values are characteristic of strong resilience. \emph{Negative} correlation suggests that liquidity gets removed/added in tandem with market orders, which would be characteristic of book ``fading'', cf.~the event-by-event plot in Figure~\ref{fig:lob-dynamics}.
Rapidly declining $\rho^j$ could then be an indication of the cumulative effect of persistent and aggressive buying/selling that leads market makers to adjust their reference prices to account for added inventory risk. 

A major challenge is to operationalize $\rho^j$ given the asynchronous nature of the entered orders. Using the aggregated bucketed quantities $VL_k, VM_k$ is not convenient given the relatively large buckets we consider---to compute statistical correlation one needs time-series of at least a few dozen terms, while the episodes observed in Figure~\ref{fig:lob-dynamics} are much more fleeting. One attempt we made was to define the correlation on either side of the LOB as
	\begin{equation}\label{cor}
	\rho_t^{j}:=\corr(VM^j_{[t-s,t]}, VL_{[t-s,t]}^j),
	\end{equation}
computing empirical correlation over a sliding window of $s$ physical time using smaller time-based buckets. Specifically we tried  $s=1.5$ hours with time buckets of $30$ seconds. However the resulting $(\rho^j_t)$ were not very stable in time and the statistical significance of the resulting predictor was too low to be included in any of the models of Section~\ref{sec:liquidity}. Another challenge is normalizing $\rho$ across assets. For tickers with deeper queues, such as MSFT, the short-term behavior of limit flows is less tied to price formation, so that $VL^j$ and $VM^j$ are less coupled, leading to lower respective correlations $|\rho^j|$.

\begin{remark}
A related mechanism to quantify  the co-movement of market and limit orders for the purposes of explaining book liquidity are so-called \textit{toxicity} indicators. Toxicity is linked to adverse selection: in a toxic environment market-makers withdraw due to orders coming from ``informed'' traders who possess better knowledge of the future asset price. In that direction, we may mention the VPIN metric \cite{easley2012flow} which can be operationalized as  the average of the absolute trade imbalance across the $\ell$ most recent volume buckets,
	\begin{equation}\label{vpinch3}
	|\overline{TI}|_{k-\ell:k} :=\frac{1}{\ell} \sum_{i=1}^{\ell} |TI_{k-i}|,
	\end{equation}
Large values of $|\overline{TI}|_{k-\ell:k}$, are supposed to lead to less liquidity and more volatility.
Another toxicity measure described in \cite{CarmonaWebster14} is constructed from the event-based correlation between market flow and price change based on the previous $200$ market order arrivals:
	\begin{equation}\label{cor2}
	\rho^{Tox}(\tau_k) :=\corr(\Delta P_{i-200:i}, O^M_{i-200:i}).
	\end{equation}
We also tried to include these in our regressions, but preliminary results suggested that they had weak predictive power (this is consistent with the feature that all lagged variables were poor predictors relative to contemporaneous covariates). It remains an open question how to construct a relevant volume-bucketed toxicity measure.
\end{remark}

\section{Conclusion}\label{sec:conclude}

 In this article we have investigated the meso-scopic phenomenology of price formation and liquidity provision/consumption. The documented stylized facts are crucial for algorithmic scheduling of child trades, where the LOB state is used as a dynamic input rather than a static parameter. Our central take-away is that at the minute-scale the distinct order flows, both of market and limit type,  are the primary predictors of price moves. We have also observed that top-level depth and the associated book imbalance are insufficient for forecasting price impact; rather deeper book characteristics, such as $PI$ and $S$, are better suited.

Our results pave the way for further research directions. First, we stress that it is imperative to develop models for limit flows in parallel with the market flow. This is a challenging task as one must consider the book asymmetry (interaction between market and limit orders on the same side of the book), cross-effects (interactions between market buys and sells, and between limit orders on the two sides, e.g.~via book imbalance) and interaction between limit Additions and Cancellations. Second,
 our approach offers a way to disentangle price evolution into price trend (impact of market executions), book resilience(impact of limit flows) and liquidity (residual effects that drive price fluctuations beyond their expected level). This gives a starting point for quantifying LOB dynamics, for example to obtain a dynamic definition of liquidity regimes. In analogue to the developed models for individual order arrivals, we need probabilistic description for the evolution of, say, $(PI_k)$ or $SL_k$.  Third, the unique meso-scopic phenomena raise new questions on connecting the event-based market microstructure, the time-based macro-diffusion and the volume-based bucketing proposed herein in a single multi-scale framework.

 First-generation models for execution, liquidation and market-making have been prescriptive in nature, postulating some features of the limit order books and optimizing participant behavior under those assumptions. The phenomenological investigation herein takes the opposite view, attempting to crystalize the data-driven features. For the practical application, the next step would be to develop \emph{predictive} models that can statistically forecast book characteristics going forward, and therefore inform execution algorithms.  Theoretically, we need models that move away from working with static book snapshots (e.g.~an abstract ``depth'' parameter) and towards using more robust definitions of impact. More sophisticated frameworks for capturing the interaction between market makers and observed flows, the so-called ``toxicity'' effects, are also warranted.

 By design, our analysis has been limited to the most liquid, large-tick equities. While these cover many of the most active tickers, including many blue-chip companies such as Dow Jones components, more work is needed on other ticker types. This would include liquid small-tick assets, such as AAPL or GOOG,  which might be very active, but where the spread has non-trivial dynamics and book depth has to be understood differently. One challenge for such tickers is that a nontrivial portion of the limit orders are aggressive, i.e.~placed inside the spread, which requires to adjust the meaning of $VL$.
 Another issue is that when share prices are very high, the discrete nature of order volumes becomes prominent. A different direction  for future work is analysis of large-tick but illiquid assets, where the event-by-event scale might still be relevant even on the minutes-scale.

\bibliography{mml}
\bibliographystyle{siam}

\appendix
\renewcommand{\thesubsection}{\Alph{subsection}}
\section*{Appendices}
\subsection{Volume Bucketing Parameters}
\begin{table}[h!]
$$  \begin{array}{c|rcrrcrrcr} \hline
 & \multicolumn{3}{ c| }{ 0.25\% \text{ADV}} & \multicolumn{3}{ c| }{ 1\% \text{ADV} } & \multicolumn{3}{ c }{2\% \text{ADV}} \\
 \text{Ticker} & V & \text{med}(\Delta \tau)  & \text{Trades} & V & \text{med} (\Delta \tau)  & \text{Trades} &
 V & \text{med}(\Delta \tau)  & \text{Trades} \\ \hline \hline
  \text{BBBY}  & 1.5K & 32.9 & 7.8 & 6K & 161 & 31.2 &  12K & 334 & 62.5 \\
  \text{INTC}  & 9K & 21.4 & 5.2 & 36K & 149 & 20.7 & 72K & 324 & 41.4 \\
  \text{MSFT}  & 25K & 28.0 & 14.6 & 100K & 159 & 58.2 & 200K & 341 & 116.5 \\
  \text{NTAP}  & 2K & 36.6 & 7.9 & 8K & 188 & 31.7 & 16K & 393 & 63.5 \\
  \text{ORCL}  & 5K & 31.6 & 6.0 &  20K & 170 & 23.9 & 40K & 357 & 47.8 \\
  \text{TEVA}  & 3K & 31.1 & 10.2 & 12K & 158 & 40.9 & 24K & 337 & 81.9 \\
   \hline\hline
\end{array}
$$
\caption{Summary statistics for the volume-based bucketing. We list the size $V$ of each bucket (in 000's executed shares), median duration $med(\tau_{k+1}-\tau_k)$ in seconds, and average number of  market trades per slice. \label{tbl:bucketing} }
\end{table}

\newgeometry{margin=0.9cm}
\begin{landscape}
\pagestyle{empty}

\subsection{Predictor Importance}
\addtolength{\arraycolsep}{-2pt}
\begin{table}[h!] \begin{center}
 $$\begin{array}{|r|cccccc|cccccc|cccccc|}
\hline
& \multicolumn{6}{ c| }{ \text{Price Change } \Delta P} & \multicolumn{6}{ c| }{\text{Ask Scarce Liquidity } SL^A} &\multicolumn{6}{ c| }{ \text{Bid Scarce Liquidity } SL^B} \\
 \text{Predictor} & \text{\small BBBY} & \text{\small  INTC} & \text{\small  MSFT} & \text{\small  ORCL} & \text{\small  NTAP} & \text{\small TEVA} &
 \text{\small BBBY} & \text{\small  INTC} & \text{\small  MSFT} & \text{\small  ORCL} & \text{\small  NTAP} & \text{\small TEVA} &
 \text{\small BBBY} & \text{\small  INTC} & \text{\small  MSFT} & \text{\small  ORCL} & \text{\small  NTAP} & \text{\small TEVA} \\ \hline \hline
TI_k     &      ** & ** & ** & ** & ** & **     &   ** & * & ** & ** & ** & * &   ** & ** & ** & ** & ** & ** \\
TI_{k-3:k-1} &         - & - & - & - & -  &  -       &   * & * & * & * & * &  *     &   - & * & * & * & - &*\\
TI_{k-1} &         - & - & - & - & -  &  -       &   * & * & * & * & - &  *     &   - & * & * & * & * &-\\
VL^a_k   &      *  & ** & * & ** & **  & **   & ** & ** & ** & ** & **  & **  &   ** & ** & ** & ** & ** & ** \\
VL^b_k   &        * & ** & * & ** & ** &  **  &  ** & ** & ** & ** & ** &  ** &   ** & ** & ** & ** & ** & ** \\
VL^a_{k-1} &         - & - & - & - & -  & -     &   * & - & - & - & - & -      &   * & * & * & - & * &* \\
VL^b_{k-1} &         - & - & - & -  & -  & -    &   * & * & *  &*  & - & *     &   * & - & - & * & - &- \\
PC^a_k &         * & * & * & * & *  & *       &   ** & ** & ** & ** & ** & ** &   * & * & * & ** & ** & *\\
PC^b_k  &        * & * & * & * & *  & *       &   * & * & * & ** & ** & *     &  ** & ** & ** & ** & ** & **\\
PI^a_k   &      * & * & * & *  & * & *        &   ** & ** & ** & ** & ** & ** &  * & ** & * & * & ** & * \\
PI^b_k   &      * & * &  *  & * & * & *    &   * & **  & ** & ** & ** & *     &  ** & ** & ** & ** & ** & ** \\
S^a_k   &      * & * & * & * & *  & -        &  ** & ** & ** & ** & ** & **   &   ** & ** & * & * & ** & * \\
S^b_k   &      * & * & * & * & *  & -        &  * & ** & ** & ** & ** & *    &   ** & ** & ** & ** & ** & **\\
D^a_1  &       - & - & - & - & -  &  -     &  * & * & * & * & * & *         &   - & * & * &  * & * & * \\
D^b_1  &      - & - & - & - & *  &  -      &  * & ** & * & * & ** & *      &   * & * & * & * & * & * \\
D_2^a  &     * & * & * & * & *  &  *       &   ** & ** & ** & ** & ** & **  &   ** & ** & ** & ** & ** & * \\
D_2^b  &     * & * & * & * & *  &  *       &   ** & ** & ** & ** & ** & ** &   ** & ** & ** & ** &  ** & **\\
BI_{k-3:k} &        -& - & - & - & *  & -        &   * & * & * & * & ** & *     &   * & * & * & * & ** & * \\
\tau_k   &       * & - & - & * & *  & -       &   ** & ** & * & ** & ** &  **&   * & * &*  & ** & ** & *\\
\tau_k-\tau_{k-1} &  - & - & - & - &-   & -     &   * & * & * & ** & ** & **   &   * & * & * & * & ** & ** \\
P_{k}-P_{k-20} &         - & - & - & - & -  & -       &   * & - & * & - & * &  -     &   - & - & - & - & * &- \\
 \hline \hline \end{array}$$
\begin{minipage}{\textwidth}
\caption{ Significant predictors for price change $\Delta P$ and scarce liquidity  $SL^j$ using a Random Forest model. We use a RF with $500$ trees, and the mean decrease in accuracy variable importance metric reported in \texttt{randomForest::importance} function.  Highly significant (**) predictors are those with relative importance above 0.4; significant (*) predictors have relative importance in $[0.15,0.4]$; the rest are labeled as not significant (-). Results are based on 1\% ADV bucketing. Predictors with subscripts other than $k$ are (averaged) lagged values over the indicated previous buckets. \label{tbl:rf-predictors} } \end{minipage} \end{center}
\end{table}
\end{landscape}

\restoregeometry

\begin{figure}
      PLOT OMITTED FROM ARXIV DUE TO SIZE
      \caption{Pairs plot for main predictors of scarce liquidity. Blue points indicate buckets with Ask-side scarce liquidity $SL^A=1$, and green are buckets with $SL^B=1$. Plot for NTAP and $V=8000$ (1\% ADV). Lower-triangle entries indicate the correlation between the predictor pairs over the entire sample period. \label{fig:pairs}}
    \end{figure}

	\begin{table}[ht]
		\begin{center}$$
			\begin{array}{c|rrrr|rrr}
			\hline \hline \vspace*{2pt}
& & \multicolumn{3}{ c }{ SL^A } & \multicolumn{3}{ |c }{ SL^B }\\
	 & & \text{Low } & \text{Med } & \text{High } & \text{Low } & \text{Med } & \text{High} \\ \hline \hline
 \multirow{ 3}{*}{\text{BBBY}} & & \multicolumn{3}{ c }{ \text{AUC}=0.938} & \multicolumn{3}{ |c }{  \text{ AUC}=0.942}\\
   &No SL	&		 84.73 &  10.71 &   0.40 &   84.28 & 10.59  &    0.37 \\
   & SL  &  0.77 &  2.76  &  0.63 &    0.68 &  3.02   &  1.05 \\ \hline		
 \multirow{ 3}{*}{\text{INTC}} & & \multicolumn{3}{ c }{ \text{   AUC}=0.928} & \multicolumn{3}{ |c }{  \text{  AUC}=0.925}\\
   &No SL &   86.67& 7.98  &  0.24 &  86.55 & 8.61   & 0.55 \\
& SL   &  1.34& 3.04  &  0.73 &   0.97 & 2.47  &  0.85 \\ \hline
\multirow{ 3}{*}{\text{MSFT}} & & \multicolumn{3}{ c }{ \text{  AUC}=0.946} & \multicolumn{3}{ |c }{ \text{   AUC}=0.963}\\
 &No  SL & 86.62& 8.33  &  0.24 &   87.96 &7.47  &  0.18 \\
&SL   & 0.77 &2.36  &  1.68 &    0.49 &2.70  &  1.19 \\ \hline
\multirow{ 3}{*}{\text{NTAP}} & & \multicolumn{3}{ c }{ \text{ AUC}=0.958} & \multicolumn{3}{ |c }{  \text{ AUC}=0.961}\\
 &        No  SL & 86.11 & 8.74 &   0.34 &   86.62& 8.12  &  0.31 \\
         &  SL  &  0.72 & 2.97 &   1.13 &   0.51& 3.11  &  1.33 \\ 	 \hline	
\multirow{ 3}{*}{\text{ORCL}} & & \multicolumn{3}{ c }{ \text{ AUC}=0.959} & \multicolumn{3}{ |c }{  \text{AUC}=0.967}\\
	&	  No SL &  86.51 & 8.54 &   0.33 &  87.62 & 7.72  &  0.26 \\
& SL   & 0.56 & 2.69  &  1.38 &    0.43 & 3.02 &   0.95  \\ \hline
\multirow{ 3}{*}{\text{TEVA}}& & \multicolumn{3}{ c }{\text{ AUC}=0.934} & \multicolumn{3}{ |c }{  \text{ AUC}=0.945}\\
  &           No SL &  85.95 & 9.45 &    0.24  &   86.67 & 8.37 &    0.27 \\
         &   SL  &   0.93  & 2.67  &    0.75  &   0.78 & 2.85  &   1.05 \\ \hline
			\end{array}$$
		\end{center}
		\caption{Goodness-of-fit for predicting Scarce Liquidity. We list observed percentages of scarce liquidity $SL$ (rows) grouped by predicted model response (columns). Specifically, let $\widehat{\pi}^j_k$ be the predicted probability of $SL^j_k$ as obtained from individual tree votes in a RF logistic regression model \eqref{logistic} with $500$ trees. Then $Low = \{ k: \widehat{\pi}_k \le 0.1\}$, $Med = \{ \widehat{\pi} \in (0.1, 0.5)\}$, $High = \{ \widehat{\pi} \ge 0.5\}$.
AUC is the area under the receiver operator curve, computed using a hold-out test set: 2/3 of data was randomly selected for training and the rest for above testing.  As a result, the actual frequency of $SL^j$ in the test sets varies across assets in the range 4\%--5\%.
All results are based on 1\% ADV buckets.
			\label{tab4}}
	\end{table}

\pagestyle{empty}
\newgeometry{margin=0.9cm}

\begin{landscape}
\subsection{Relationship between Market and Limit Order Flows}
	\begin{figure}[!ht]
		\centering
     \begin{tabular}{cc}
		{\includegraphics[height=1.85in, trim=0in 0.3in 0in 0.4in]{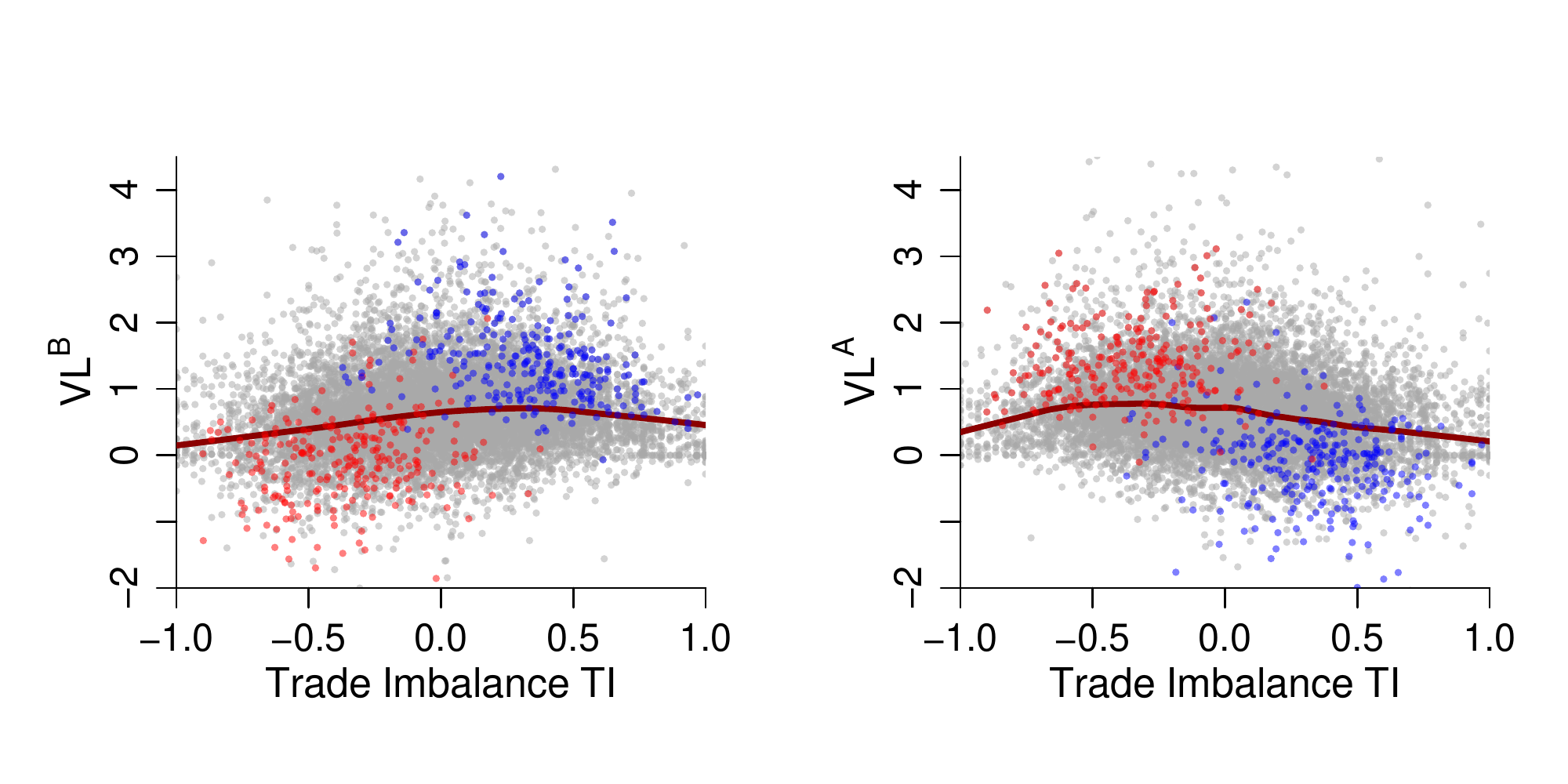}} &
        \includegraphics[height=1.85in, trim=0.1in 0.1in 0.3in 0.1in]{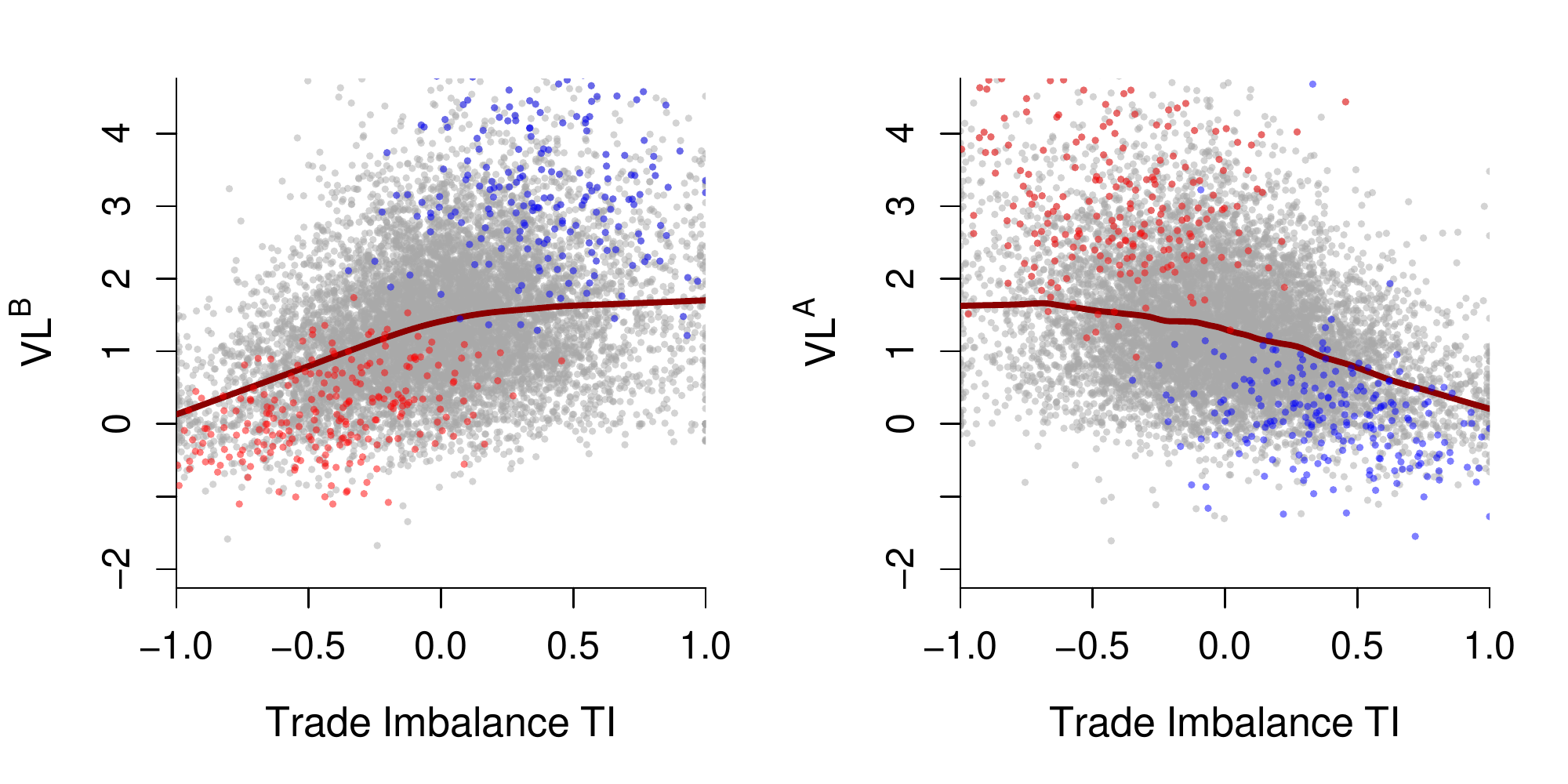}  \\
         BBBY $V=6,000$ & ORCL $V=20,000$ \\
          \includegraphics[height=1.85in,trim=0.4in 0.3in 0.25in 0in]{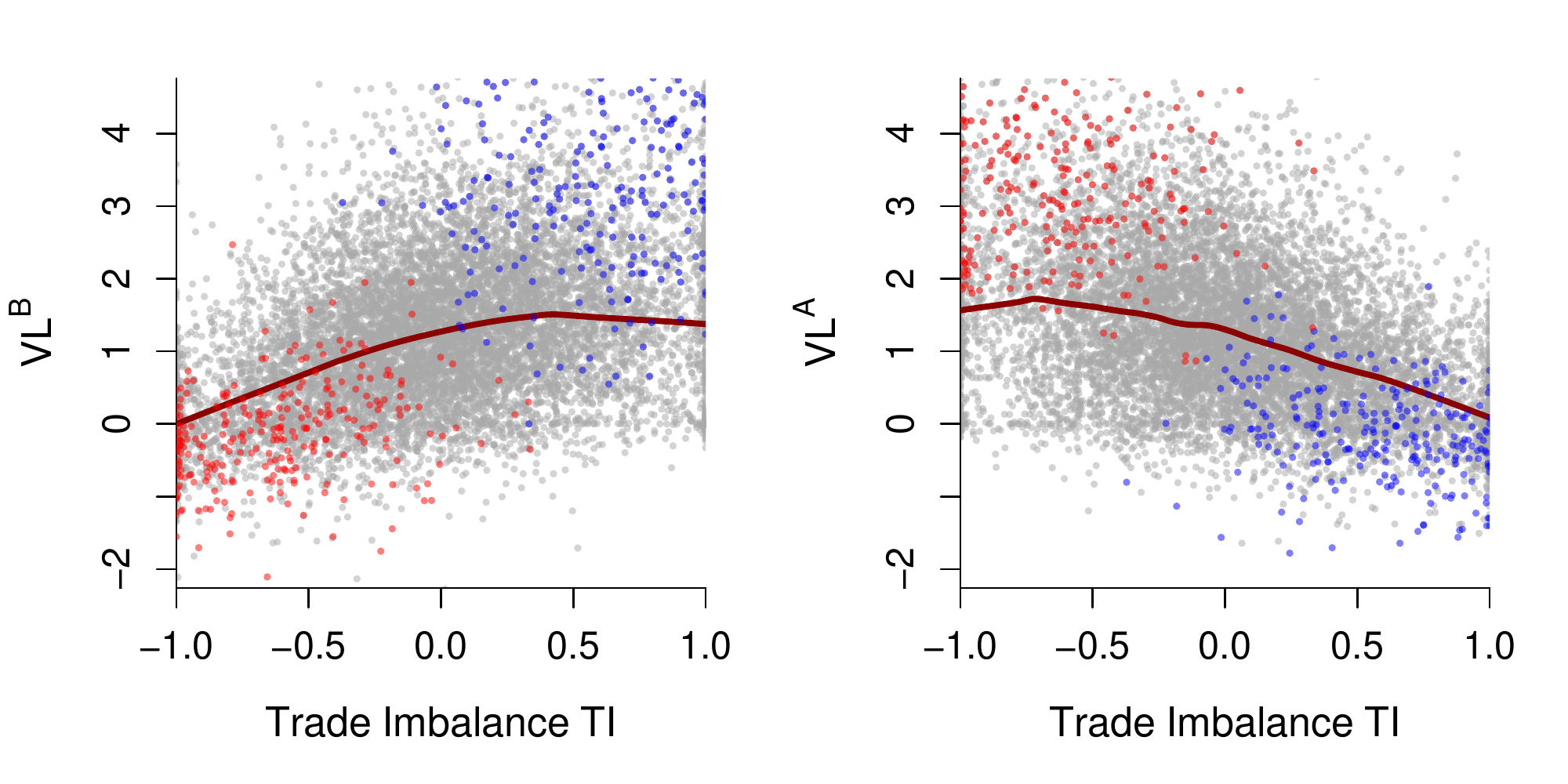}  &
        \includegraphics[height=1.85in, trim=0.1in 0.1in 0.3in 0in]{VLvsTI-ntap8} \\
        INTC $V=36,000$ & NTAP $V=8,000$ \\
        {\includegraphics[height=1.85in, trim=0.4in 0.4in 0.25in 0in]{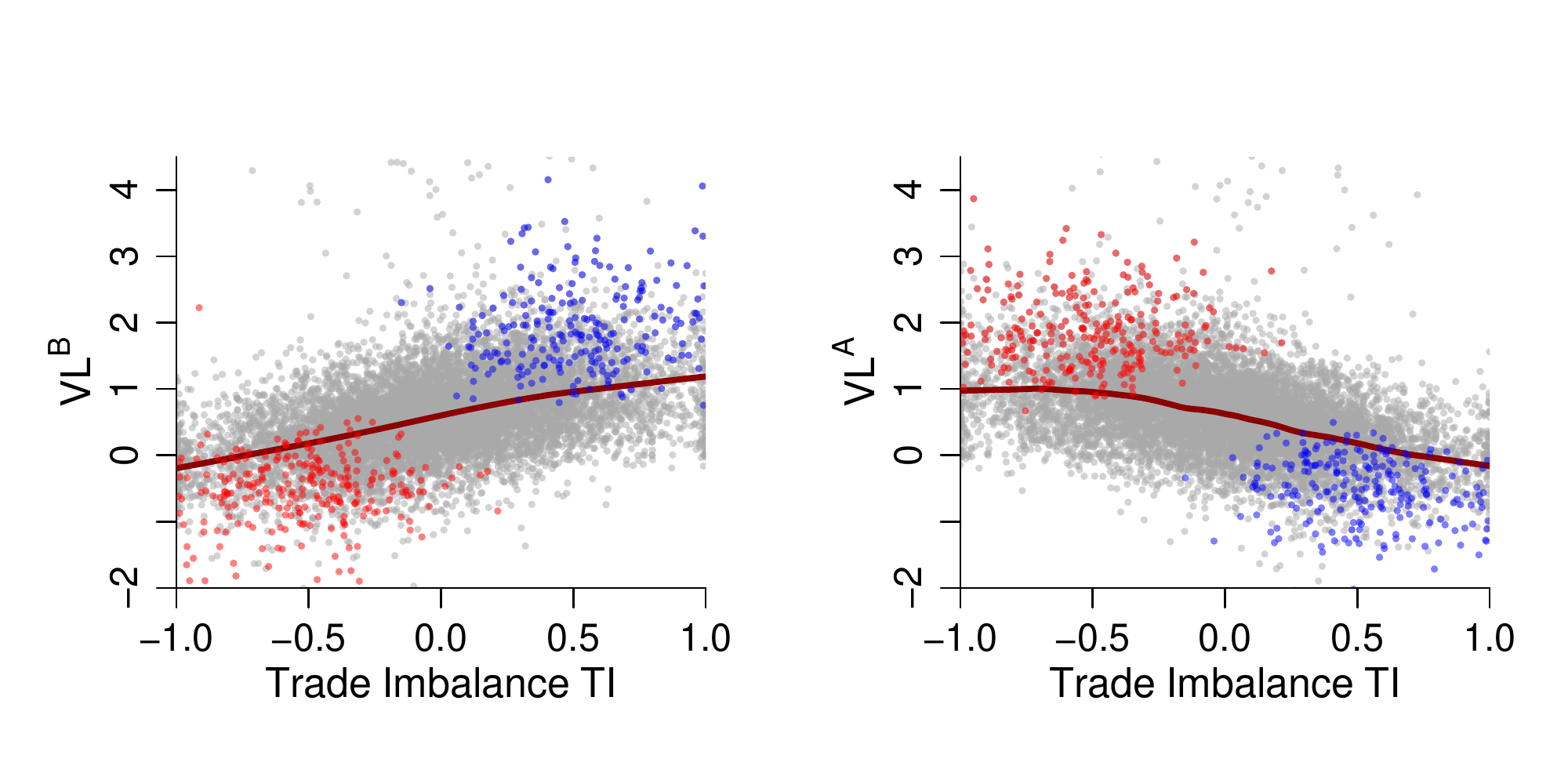}}  &
        \includegraphics[height=1.85in, trim=0.1in 0.2in 0.3in 0in]{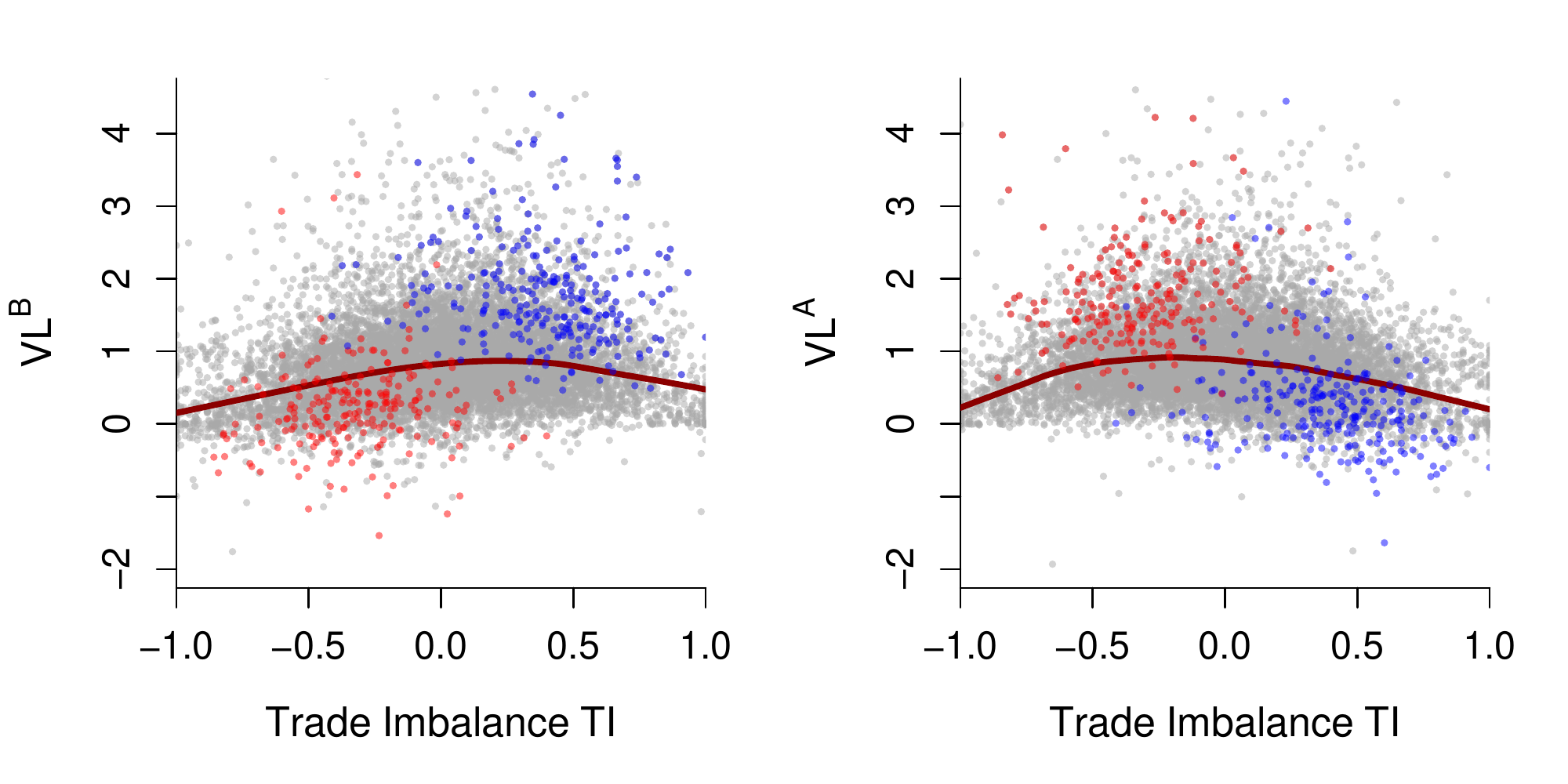} \\
        MSFT $V=100,000$ & TEVA $V=12,000$ \\
        \end{tabular}
		\begin{minipage}{\textwidth}
\caption{Net limit order flow at the best bid (left panels) and best ask (right panels) plotted against the market order imbalance $TI$. The $y$-axis is normalized via $VL^j/V$. Blue (red) points indicate buckets with large price moves: $\Delta P_k$ of more (less) than $ \pm 1.645 \cdot StDev( \Delta P)$. The solid line shows the GAM fit for $g^j(TI)= \E[ VL^j | TI]$, cf.~Figure~\ref{fig:VLvsTI}. \label{fig:allVLvsTI}} \end{minipage}
	\end{figure}
\end{landscape}

\end{document}